# Classical and Bayesian Analyses of a Mixture of Exponential and Lomax Distributions

Maqsood Ali[a*], Abdul Haq[b], Muhammad Aslam[c]

[a]Ludwig-Maximilians-Universität, Munich, Germany

[b]Quaid-i-Azam University, Islamabad, Pakistan

[c]Riphah International University, Islamabad, Pakistan

**Abstract:** The exponential and the Lomax distributions are widely used in life testing experiments in mixture models. A mixture model of exponential distribution and Lomax distribution is proposed. Parameters of the proposed model are estimated using classical and Bayesian procedures under type-I right censoring. Expressions for Bayes estimators are derived assuming noninformative (uniform and Jeffreys) priors under symmetric and asymmetric loss functions. Posterior predictive distributions of a future observation are derived and predictive estimates are obtained. Extensive Monte Carlo simulations are carried out to investigate performance of the estimators in terms of sample sizes, censoring times and mixing proportions. The analysis of mixture model is carried out using a data set of lifetime of transmitter receivers. Interesting properties of estimators are observed and discussed.

[*] Corresponding author: E-mail: maqsoodfsd@outlook.com

# 1. Introduction

The importance of finite mixture models in statistical analysis of data can be judged by ever-increasing rate at which articles on applications of mixture model appear in literature. Because of the flexibility of mixture models, these models are being increasingly used in a variety of applications e.g., reliability and survival analysis, medical diagnosis and prognosis, ecology, fishery, biology, astronomy, quality control and econometrics. A variety of applications exist in mixture models e.g., beauty-contest data is analyzed using the finite mixture model in Bosch-Domènech et al. (2010). Joffe (1964) fits the frequency distribution of dust particles in mines using mixture models. Everitt and Hand (1981), Titterington et al. (1985), McLachlan and Peel (2000) and Erişoğlu et al. (2011)) can be consulted for more applications.

In situations where the complete data sets are unavailable, we focus on the more difficult situations of incomplete data sets. Consider an object is put on life testing experiment and the complete life length cannot be determined or unknown, the object is then considered a censored object. Such a censoring is called right censoring of type-I, with fixed test termination time.

Mixtures of lifetime distributions such as Weibull-gamma, Weibull-exponential and exponential-gamma are proposed in Erişoğlu et al. (2011). Saleem et al. (2009) proposed a two-component mixture of power function distribution and addressed the problem of estimation through Bayesian approach. Al-Hussaini et al. (2001) used a two-component mixture model of Lomax distributions to obtain the Bayesian predictive bounds. The Lomax distribution is used for stochastic modeling of decreasing failure rate life components in accelerated life testing analyses. For example, the scale parameter of Lomax distribution is formed as a function of stress levels in Hassan and Al-Ghamdi (2009). It makes the motivation to propose a mixture of two different lifetime distributions, i.e., exponential and Lomax distributions. The proposed mixture model is useful in modeling the lifetimes of heterogeneous population which will be composed of exponential and Lomax distributions.

The mixture model is proposed in Section 2 and Section 3 deals with formation of likelihood function and construction of system of equations to find maximum likelihood (ML) estimates. Bayesian analysis of proposed model assuming uniform and Jeffreys priors is carried out in Section 4. Loss functions and importance sampling procedure used in this study are also highlighted in this section. The expressions for Bayes estimates and posterior variances are obtained. Bayes estimates are computed using importance sampling procedure too. Section 5 presents the posterior predictive distributions and Bayes point predictors. Predictive bounds are also constructed for future observation. Simulations are performed in Section 6 and analysis using real life data set is addressed in Section 7. This study finishes off by drawing some conclusions in Section 8.

## 2. The Model

In a mixture model, different probability distributions are mixed together. Probability distributions may be mixed in terms of different components or the same components having different parameters. A mixture model having two components with mixing proportion $p$ may be defined as

$$f(x) = pf_1(x) + (1-p)f_2(x), \qquad 0 < p < 1. \tag{2.1}$$

The probability density function of an exponential random variable $X$ is

$$f(x; \theta_1) = \theta_1 \exp(-\theta_1 x), \qquad x \geq 0,\ \theta_1 > 0, \tag{2.2}$$

where $\theta_1$ is the scale parameter of distribution. The distribution function of (2.2) is $F(x) = 1 - \exp(-\theta_1 x)$. The Lomax distribution of a random variable $X$ having shape parameter $\theta_2$ and scale parameter $\delta$ is

$$f(x; \theta_2, \delta) = \theta_2 \delta^{\theta_2} (x + \delta)^{-(\theta_2 + 1)}, \qquad x \geq 0,\ \theta_2, \delta > 0. \tag{2.3}$$

The distribution function of (2.3) is $F(x) = 1 - \left(1 + \dfrac{x}{\delta}\right)^{-\theta_2}$. It is assumed that the population is heterogeneous and contains the data belonging independently from (2.2) and (2.3). Then the mixture model for such a population by following (2.1) is

$$f(x; \theta_1, \theta_2, \delta, p) = p\theta_1 \exp(-\theta_1 x) + (1-p)\theta_2 \delta^{\theta_2}(x + \delta)^{-(\theta_2 + 1)}. \tag{2.4}$$

Here, $\delta = 1$ is assumed known and the parameters of interest are $\theta_1$, $\theta_2$ and $p$. The distribution function of (2.4) is

$$F(x) = pF_1(x) + (1-p)F_2(x),$$

$$F(x; \theta_1, \theta_2, \delta, p) = 1 - p\exp(-\theta_1 x) - (1-p)\left(1 + \dfrac{x}{\delta}\right)^{-\theta_2}, \tag{2.5}$$

where $F_1(x)$ and $F_2(x)$ are the distribution functions corresponding to (2.2) and (2.3) respectively.

## 3. The Likelihood Function

Suppose $n$ units are put in a life testing experiment and let $r$ units are failed up to the time $T$. And the remaining $n - r$ units are still functioning. Out of failed units, let $r_1$ and $r_2$ units belong to first and second subpopulations such that $r_1 + r_2 = r$. Let $x_{ij}$, $0 < x_{ij} \leq T$, denotes the failure time of $j^{th}$ unit belonging to $i^{th}$ subpopulation, $i = 1, 2$, $j = 1, 2, 3, \cdots, r_i$. Our interest is to estimate the parameters of proposed mixture model (2.4) based on a sample censored at a prefixed time $T$. The likelihood function of mixture model (2.4) for the observed random sample $\mathbf{x} = (x_{11}, x_{12}, \cdots, x_{1r_1}, x_{21}, x_{22}, \cdots, x_{2r_2})$ is defined as

$$L(\boldsymbol{\theta} \mid \mathbf{x}) \propto \left\{\prod_{j=1}^{r_1} p f_1(x_{1j})\right\} \left\{\prod_{j=1}^{r_2} (1-p) f_2(x_{2j})\right\} \{1 - F(t)\}^{n-r},$$

$$L(\boldsymbol{\theta} \mid \mathbf{x}) \propto \left\{\prod_{j=1}^{r_1} p \theta_1 \exp(-\theta_1 x_{1j})\right\} \left\{\prod_{j=1}^{r_2} (1-p) \theta_2 \delta^{\theta_2} (x_{2j} + \delta)^{-(\theta_2 + 1)}\right\}$$

$$\times \left\{\left(p \exp(-\theta_1 t) + (1-p)\left(1 + \frac{t}{\delta}\right)^{-\theta_2}\right)^{n-r}\right\}, \quad (3.1)$$

where $\boldsymbol{\theta} = (\theta_1, \theta_2, p)$. After some simplification, it can be expressed in the following form

$$L(\boldsymbol{\theta} \mid \mathbf{x}) \propto \theta_1^{r_1} \theta_2^{r_2} \sum_{k=0}^{n-r} \binom{n-r}{k} \exp\left\{-\theta_1 \left(\sum_{j=1}^{r_1} x_{1j} + tk\right)\right\}$$

$$\times p^{k+r_1} (1-p)^{n-k-r_1} \exp\left\{-\theta_2 \left(\sum_{j=1}^{r_2} \ln\left(1 + \frac{x_{2j}}{\delta}\right) + (n-r-k) \ln\left(1 + \frac{t}{\delta}\right)\right)\right\}. \quad (3.2)$$

### 3.1. Maximum Likelihood Estimates

The maximum likelihood (ML) estimates of parameters $\theta_1$, $\theta_2$ and $p$ are obtained by solving the non-linear system of equations given by (3.3)–(3.5). The system of equations is constructed by partially differentiating the natural logarithm of (3.2) with respect to $\theta_1$, $\theta_2$ and $p$ respectively. Consider $l = \ln\{L(\boldsymbol{\theta} \mid \mathbf{x})\}$, for simplicity, then

$$l = \ln(\eta) + r_1 \ln(p) + r_1 \ln(\theta_1) - \theta_1 \sum_{j=1}^{r_1} x_{1j} + r_2 \ln(1-p) + r_2 \ln(\theta_2) + r_2 \theta_2 \ln(\delta)$$

$$- (\theta_2 + 1) \sum_{j=1}^{r_2} \ln(x_{2j} + 1) + (n-r) \ln\left\{p \exp(-\theta_1 t) + (1-p)\left(1 + \frac{t}{\delta}\right)^{-\theta_2}\right\},$$

where $\eta$ is constant of proportionality in (3.2) and $\ln(\cdot)$ is the natural logarithm.

$$\frac{\partial l}{\partial \theta_1} = \frac{r_1}{\theta_1} - \sum_{j=1}^{r_1} x_{1j} - \frac{1}{C}(n-r) p t \exp(-\theta_1 t), \quad (3.3)$$

$$\frac{\partial l}{\partial \theta_2} = \frac{r_2}{\theta_2} - \sum_{j=1}^{r_2} \ln\left(\frac{x_{2j}}{\delta} + 1\right) - \frac{1}{C}(n-r)(1-p)\left(1 + \frac{t}{\delta}\right)^{-\theta_2} \ln\left(1 + \frac{t}{\delta}\right) \quad (3.4)$$

and $\quad \dfrac{\partial l}{\partial p} = \dfrac{r_1}{p} - \dfrac{r_2}{1-p} + \dfrac{1}{C}(n-r)\left\{\exp(-\theta_1 t) - \left(1 + \dfrac{t}{\delta}\right)^{-\theta_2}\right\}, \quad (3.5)$

where $C = p\exp(-\theta_1 t) + (1-p)\left(1+\dfrac{t}{\delta}\right)^{-\theta_2}$. The numerical methods are used to obtain the ML estimates of $\theta_1$, $\theta_2$ and $p$.

### 3.2. Variances of ML Estimates

The main diagonal of the inverted Fisher's information matrix provides the variances of ML estimates. Unfortunately, the exact expressions for the expectations are not easy to find. Therefore, numerical methods can be used to approximate the expressions for the expectations. The information matrix is given by

$$I(\boldsymbol{\theta}) = -E\begin{bmatrix} \dfrac{\partial^2 l}{\partial \theta_1^2} & \dfrac{\partial^2 l}{\partial \theta_1 \partial \theta_2} & \dfrac{\partial^2 l}{\partial \theta_1 \partial p} \\ \dfrac{\partial^2 l}{\partial \theta_2 \partial \theta_1} & \dfrac{\partial^2 l}{\partial \theta_2^2} & \dfrac{\partial^2 l}{\partial \theta_2 \partial p} \\ \dfrac{\partial^2 l}{\partial p \partial \theta_1} & \dfrac{\partial^2 l}{\partial p \partial \theta_2} & \dfrac{\partial^2 l}{\partial p^2} \end{bmatrix}, \qquad (3.6)$$

where 
$$\dfrac{\partial^2 l}{\partial \theta_1^2} = -\dfrac{r_1}{\theta_1^2} + \dfrac{1}{C}(n-r)pt^2 \exp(-\theta_1 t) + \dfrac{1}{C^2}(n-r)p^2 t^2 \exp(-2\theta_1 t), \qquad (3.7)$$

$$\dfrac{\partial^2 l}{\partial \theta_2^2} = -\dfrac{r_2}{\theta_2^2} - \dfrac{1}{C^2}(n-r)(1-p)^2 \left(1+\dfrac{t}{\delta}\right)^{-2\theta_2 t} \left\{\ln\left(1+\dfrac{t}{\delta}\right)\right\}^2$$
$$+ \dfrac{1}{C}(n-r)(1-p)\left(1+\dfrac{t}{\delta}\right)^{-\theta_2}\left[(\ln \delta)^2 - 2\ln\delta \ln(\delta+t) + \{\ln(\delta+t)\}^2\right], \quad (3.8)$$

$$\dfrac{\partial^2 l}{\partial p^2} = -\dfrac{r_1}{p^2} - \dfrac{r_2}{(1-p)^2} - \dfrac{1}{C^2}(n-r)\left\{\exp(-\theta_1 t) - \left(1+\dfrac{t}{\delta}\right)^{-\theta_2}\right\}^2, \qquad (3.9)$$

$$\dfrac{\partial^2 l}{\partial \theta_1 \partial \theta_2} = \dfrac{\partial^2 l}{\partial \theta_2 \partial \theta_1} = -\dfrac{1}{C^2}(n-r)p(1-p)t\exp(-\theta_1 t)\left(1+\dfrac{t}{\delta}\right)^{-\theta_2}\ln\left(1+\dfrac{t}{\delta}\right), \qquad (3.10)$$

$$\dfrac{\partial^2 l}{\partial p \partial \theta_1} = \dfrac{\partial^2 l}{\partial \theta_1 \partial p} = \dfrac{1}{C}(n-r)t\exp(-\theta_1 t)\left[-1 + \dfrac{1}{C}\left\{\exp(-\theta_1 t) - \left(1+\dfrac{t}{\delta}\right)^{-\theta_2}\right\}\right] \qquad (3.11)$$

and 
$$\dfrac{\partial^2 l}{\partial p \partial \theta_2} = \dfrac{\partial^2 l}{\partial \theta_2 \partial p} = \dfrac{1}{C^2}(n-r)(1-p)\left(1+\dfrac{t}{\delta}\right)^{-\theta_2}\ln\left(1+\dfrac{t}{\delta}\right)\left\{\exp(-\theta_1 t) - \left(1+\dfrac{t}{\delta}\right)^{-\theta_2}\right\}$$
$$+ \dfrac{1}{C}(n-r)\left(1+\dfrac{t}{\delta}\right)^{-\theta_2}\ln\left(1+\dfrac{t}{\delta}\right). \qquad (3.12)$$

## 4. Bayesian Estimation

This Section provides the expressions of Bayes estimators under squared error loss function (SELF) and general entropy loss function (GELF), and posterior variances using uniform and Jeffreys priors. Below described importance sampling technique is also used to obtain approximate Bayes estimators.

## 4.1. Bayes Estimators under Loss Functions

A loss is associated with an estimate being different from either a true or a desired value. We use two loss functions (SELF and GELF) for the estimation of Bayes estimators. The SELF is symmetric in nature and is defined as

SELF: $\quad L(\theta, \hat{\theta}) = (\hat{\theta} - \theta)^2$.

The mean of the posterior distribution is Bayes estimator under SELF. i.e., $\hat{\theta}_S = E_{\theta|\mathbf{x}}(\theta)$. In some situations, asymmetric loss functions have been functional (see Zellner (1986) and Calabria and Pulcini (1996)). The asymmetric GELF is defined as

GELF: $\quad L(\theta, \hat{\theta}) = \omega \left\{ \left(\frac{\hat{\theta}}{\theta}\right)^c - c \ln\left(\frac{\hat{\theta}}{\theta}\right) - 1 \right\}, \quad \omega > 0, \, c \neq 0,$

It is assumed that $\omega = 1$. The constant $c$ represents the effect of over or under estimation of parameters. Take $c > 0$, if over estimation (positive error) is more serious and $c < 0$ is used if under estimation (negative error) is more serious. The Bayes estimator of parameter $\theta$ under GELF is: $\hat{\theta}_G = \left\{ E_{\theta|\mathbf{x}}(\theta^{-c}) \right\}^{-\frac{1}{c}}$.

## 4.2. Importance Sampling

It is a technique for assessing properties of a distribution under consideration when sampling is done from some other distribution(s). The basic idea of importance sampling is that certain values of input random variables can be easily sampled. So, we focus attention on locating such region in the form of a probability distribution (which is close to the distribution of interest) that encourages such certain values. Ghosh et al. (2006) provide a detailed insight to the importance sampling procedure. We use importance sampling to obtain Bayes estimators and call them approximate Bayes estimators.

## 4.3. Posterior Distribution Assuming the Uniform Priors

We assumed uniform distribution over $(0, \infty)$ as uniform priors for $\theta_1$ and $\theta_2$. And uniform prior over $(0,1)$ is taken for mixing parameter $p$. The independent joint prior distribution of $\theta_1$, $\theta_2$ and $p$ in density kernel form is

$$\pi_2(\boldsymbol{\theta}) \propto 1, \qquad \theta_1, \theta_2 > 0, \, 0 < p < 1. \qquad (4.1)$$

The joint posterior distribution is obtained by incorporating the density kernel (4.1) with the likelihood (3.2). The marginal posterior distributions are obtained by integrating out the irrelevant parameters.

### 4.3.1. Bayes Estimators under SELF Assuming the Uniform Priors

Posterior means or Bayes estimators under SELF are obtained from marginal posterior distributions and the resultant expressions are

$$\hat{\theta}_{1US} = \frac{\Gamma(r_1+2)\Gamma(r_2+1)}{H_2} \sum_{k=0}^{n-r} \binom{n-r}{k} A_{2k}^{-(r_1+2)} B_{2k}^{-(r_2+1)} \text{Beta}(k+r_1+1, n-k-r_1+1), \quad (4.2)$$

$$\hat{\theta}_{2US} = \frac{\Gamma(r_1+1)\Gamma(r_2+2)}{H_2} \sum_{k=0}^{n-r} \binom{n-r}{k} A_{2k}^{-(r_1+1)} B_{2k}^{-(r_2+2)} \text{Beta}(k+r_1+1, n-k-r_1+1) \quad (4.3)$$

$$\text{and} \quad \hat{p}_{US} = \frac{\Gamma(r_1+1)\Gamma(r_2+1)}{H_2} \sum_{k=0}^{n-r} \binom{n-r}{k} A_{2k}^{-(r_1+1)} B_{2k}^{-(r_2+1)} \text{Beta}(k+r_1+2, n-k-r_1+1). \quad (4.4)$$

Here, $\Gamma(\cdot)$ and $\text{Beta}(\cdot)$ denote the gamma and the beta functions respectively. And where

$$A_{2k} = \sum_{j=1}^{r_1} x_{1j} + tk, \quad B_{2k} = \sum_{j=1}^{r_2} \ln\left(1+\frac{x_{2j}}{\delta}\right) + (n-r-k)\ln\left(1+\frac{t}{\delta}\right) \text{ and } H_2 \text{ is given by}$$

$$H_2 = \Gamma(r_1+1)\Gamma(r_2+1) \sum_{k=0}^{n-r} \binom{n-r}{k} A_{2k}^{-(r_1+1)} B_{2k}^{-(r_2+1)} \text{Beta}(k+r_1+1, n-k-r_1+1).$$

### 4.3.2. Bayes Estimators under GELF Assuming the Uniform Priors

The Bayes estimators under GELF are obtained by taking the expectations of the marginal posterior distributions as follows

$$\hat{\theta}_{iUG} = \left\{ E_{\theta_{iUG}|\mathbf{x}}\left(\theta_{iUG}^{-c}\right) \right\}^{-\frac{1}{c}}, \quad i=1,2 \quad \text{and} \quad \hat{p}_{UG} = \left\{ E_{p_{UG}|\mathbf{x}}\left(p_{UG}^{-c}\right) \right\}^{-\frac{1}{c}}.$$

The expressions for the Bayes estimators under GELF are

$$\hat{\theta}_{1UG} = \left\{ \frac{\Gamma(r_1+1-c)\Gamma(r_2+1)}{H_2} \sum_{k=0}^{n-r} \binom{n-r}{k} A_{2k}^{-(r_1+1-c)} B_{2k}^{-(r_2+1)} \text{Beta}(k+r_1+1, n-k-r_1+1) \right\}^{-\frac{1}{c}}, \quad (4.5)$$

$$\hat{\theta}_{2UG} = \left\{ \frac{\Gamma(r_1+1)\Gamma(r_2+1-c)}{H_2} \sum_{k=0}^{n-r} \binom{n-r}{k} A_{2k}^{-(r_1+1)} B_{2k}^{-(r_2+1-c)} \text{Beta}(k+r_1+1, n-k-r_1+1) \right\}^{-\frac{1}{c}} \quad (4.6)$$

$$\text{and} \quad \hat{p}_{UG} = \left\{ \frac{\Gamma(r_1+1)\Gamma(r_2+1)}{H_2} \sum_{k=0}^{n-r} \binom{n-r}{k} A_{2k}^{-(r_1+1)} B_{2k}^{-(r_2+1)} \text{Beta}(k+r_1-c+1, n-k-r_1+1) \right\}^{-\frac{1}{c}}. \quad (4.7)$$

### 4.3.3. Bayes Estimators with Importance Sampling Procedure

The joint uniform prior of $\theta_1$, $\theta_2$ and $p$ given in (4.1) is incorporated with the likelihood (3.1), the resulting expression may be written as

$$\psi_2(\boldsymbol{\theta}|\mathbf{x}) \propto p^{r_1}(1-p)^{r_2} \theta_1^{r_1}\theta_2^{r_2} \exp\left(-\theta_1 \sum_{j=1}^{r_1} x_{1j}\right)$$

$$\times \exp\left\{-\theta_2 \sum_{j=1}^{r_2} \ln\left(1+\frac{x_{2j}}{\delta}\right)\right\} \left\{pe^{-\theta_1 t} + (1-p)\left(1+\frac{t}{\delta}\right)^{-\theta_2}\right\}^{n-r} \quad (4.8)$$

$$\psi_2(\boldsymbol{\theta}|\mathbf{x}) \propto \Gamma_P\left(r_1+1, \sum_{j=1}^{r_1} x_{1j}\right) \Gamma_P\left(r_2+1, \sum_{j=1}^{r_2} \ln\left(1+\frac{x_{2j}}{\delta}\right)\right) \text{Beta}_P(r_1+1, r_2+1) h(\boldsymbol{\theta}), \quad 4.9)$$

where $\Gamma_P(\cdot)$ and $\text{Beta}_P(\cdot)$ denote the probability density functions of gamma and beta distributions respectively. And $h(\boldsymbol{\theta})$ is given by

$$h(\boldsymbol{\theta}) = \left\{p \exp(-\theta_1 t) + (1-p)\left(1+\frac{t}{\delta}\right)^{-\theta_2}\right\}^{n-r}. \quad (4.10)$$

Bayes estimator with importance sampling (approximate Bayes estimators) can be found as

- Step 1: Generate $\theta_1 \sim \Gamma_P\left(a_1+r_1, \sum_{j=1}^{r_1} x_{1j} + b_1\right)$, $\theta_2 \sim \Gamma_P\left(a_2+r_2, \sum_{j=1}^{r_2} \ln\left(1+\frac{x_{2j}}{\delta}\right) + b_2\right)$ and

  $p \sim \text{Beta}_P(a_3+r_1, b_3+r_2)$.

- Step 2: Obtain a set of points $(\theta_{11}, \theta_{21}, p_1)$, $(\theta_{12}, \theta_{22}, p_2), \cdots, (\theta_{1M}, \theta_{2M}, p_M)$ as in Step 1.

- Step 3: The approximate Bayes estimator of $g(\boldsymbol{\theta})$ can be determined as

$$\frac{\sum_{k=1}^{M} g(\theta_{1k}, \theta_{2k}, p_k) h(\theta_{1k}, \theta_{2k}, p_k)}{\sum_{k=1}^{M} h(\theta_{1k}, \theta_{2k}, p_k)}.$$

The approximate Bayes estimators under SELF are

$$\hat{\theta}_{iISUS} = \frac{\sum_{k=1}^{1000} \theta_{ik} h(\theta_{1k}, \theta_{2k}, p_k)}{\sum_{k=1}^{1000} h(\theta_{1k}, \theta_{2k}, p_k)}, \quad i=1,2 \quad \text{and} \quad \hat{p}_{ISUS} = \frac{\sum_{k=1}^{1000} p_k h(\theta_{1k}, \theta_{2k}, p_k)}{\sum_{k=1}^{1000} h(\theta_{1k}, \theta_{2k}, p_k)}.$$

And the approximate Bayes estimators under GELF are

$$\hat{\theta}_{iISUG} = \left\{\frac{\sum_{k=1}^{1000} \theta_{ik}^{-c} h(\theta_{1k}, \theta_{2k}, p_k)}{\sum_{k=1}^{1000} h(\theta_{1k}, \theta_{2k}, p_k)}\right\}^{-\frac{1}{c}}, \quad i=1,2 \quad \text{and} \quad \hat{p}_{ISUG} = \left\{\frac{\sum_{k=1}^{1000} p_k^{-c} h(\theta_{1k}, \theta_{2k}, p_k)}{\sum_{k=1}^{1000} h(\theta_{1k}, \theta_{2k}, p_k)}\right\}^{-\frac{1}{c}}.$$

### 4.3.4. Posterior Variances Assuming the Uniform Priors

Posterior variances determine the amount of uncertainty in the parameters. The expressions for the posterior variances are obtained from marginal posterior distributions.

$$Var_2(\theta_1 | \mathbf{x}) = \frac{\Gamma(r_1+3)\Gamma(r_2+1)}{H_2} \sum_{k=0}^{n-r} \binom{n-r}{k} A_{2k}^{-(r_1+3)} B_{2k}^{-(r_2+1)} \text{Beta}(k+r_1+1, n-k-r_1+1)$$

$$-\left\{ \frac{\Gamma(r_1+2)\Gamma(r_2+1)}{H_2} \sum_{k=0}^{n-r} \binom{n-r}{k} A_{2k}^{-(r_1+2)} B_{2k}^{-(r_2+1)} \text{Beta}(k+r_1+1, n-k-r_1+1) \right\}^2, \quad (4.11)$$

$$Var_2(\theta_2 | \mathbf{x}) = \frac{\Gamma(r_1+1)\Gamma(r_2+3)}{H_2} \sum_{k=0}^{n-r} \binom{n-r}{k} A_{2k}^{-(r_1+1)} B_{2k}^{-(r_2+3)} \text{Beta}(k+r_1+1, n-k-r_1+1)$$

$$-\left\{ \frac{\Gamma(r_1+1)\Gamma(r_2+2)}{H_2} \sum_{k=0}^{n-r} \binom{n-r}{k} A_{2k}^{-(r_1+1)} B_{2k}^{-(r_2+2)} \text{Beta}(k+r_1+1, n-k-r_1+1) \right\}^2 \quad (4.12)$$

and

$$Var_2(p | \mathbf{x}) = \frac{\Gamma(r_1+1)\Gamma(r_2+1)}{H_2} \sum_{k=0}^{n-r} \binom{n-r}{k} A_{2k}^{-(r_1+1)} B_{2k}^{-(r_2+1)} \text{Beta}(k+r_1+3, n-k-r_1+1)$$

$$-\left\{ \frac{\Gamma(r_1+1)\Gamma(r_2+1)}{H_2} \sum_{k=0}^{n-r} \binom{n-r}{k} A_{2k}^{-(r_1+1)} B_{2k}^{-(r_2+1)} \text{Beta}(k+r_1+2, n-k-r_1+1) \right\}^2. \quad (4.13)$$

### 4.4. Posterior Distribution Assuming the Jeffreys Priors

Jeffreys prior is a very famous prior among noninformative priors. It is based on observed data (see Jeffreys (1961)) and extracted from Fisher's information matrix. We assume $\theta_i \propto \sqrt{|I(\theta_i)|}$, $i=1,2$, where $I(\theta_i)$ is given by $-E\left\{ \frac{\partial^2 f(x|\theta_i)}{\partial \theta_i^2} \right\}$, and $f(x|\theta_i)$ are given in (2.2) and (2.3). And prior distribution of mixing parameter $p$ is assumed a uniform distribution over $(0,1)$. The independent joint prior distribution of $\theta_1$, $\theta_2$ and $p$ in density kernel form is

$$\pi_3(\boldsymbol{\theta}) \propto \frac{1}{\theta_1 \theta_2}, \qquad \theta_1, \theta_2 > 0, \ 0 < p < 1, \quad (4.14)$$

The joint posterior distribution is obtained by incorporating the density kernel (4.14) with the likelihood (3.2). The marginal posterior distributions are obtained by integrating out the irrelevant parameters.

### 4.4.1. Bayes Estimators under SELF Assuming the Jeffreys Priors

The expressions for the Bayes estimators under SELF are

$$\hat{\theta}_{1JS} = \frac{\Gamma(r_1+1)\Gamma(r_2)}{H_3} \sum_{k=0}^{n-r} \binom{n-r}{k} A_{2k}^{-(r_1+1)} B_{2k}^{-r_2} \text{Beta}(k+r_1+1, n-k-r_1+1), \quad (4.15)$$

$$\hat{\theta}_{2JS} = \frac{\Gamma(r_1)\Gamma(r_2+1)}{H_3} \sum_{k=0}^{n-r} \binom{n-r}{k} A_{2k}^{-r_1} B_{2k}^{-(r_2+1)} \text{Beta}(k+r_1+1, n-k-r_1+1) \tag{4.16}$$

and $$\hat{p}_{JS} = \frac{\Gamma(r_1)\Gamma(r_2)}{H_3} \sum_{k=0}^{n-r} \binom{n-r}{k} A_{2k}^{-r_1} B_{2k}^{-r_2} \text{Beta}(k+r_1+2, n-k-r_1+1). \tag{4.17}$$

Here, $H_3$ is a constant and is given by

$$H_3 = \Gamma(r_1)\Gamma(r_2) \sum_{k=0}^{n-r} \binom{n-r}{k} A_{2k}^{-r_1} B_{2k}^{-r_2} \text{Beta}(k+r_1+1, n-k-r_1+1).$$

### 4.4.2. Bayes Estimators under GELF Assuming the Jeffreys Priors

The expressions for the Bayes estimators under GELF are

$$\hat{\theta}_{1JG} = \left\{ \frac{\Gamma(r_1-c)\Gamma(r_2)}{H_3} \sum_{k=0}^{n-r} \binom{n-r}{k} A_{2k}^{-(r_1-c)} B_{2k}^{-(r_2)} \text{Beta}(k+r_1+1, n-k-r_1+1) \right\}^{-\frac{1}{c}}, \tag{4.18}$$

$$\hat{\theta}_{2JG} = \left\{ \frac{\Gamma(r_1)\Gamma(r_2-c)}{H_3} \sum_{k=0}^{n-r} \binom{n-r}{k} A_{2k}^{-(r_1)} B_{2k}^{-(r_2-c)} \text{Beta}(k+r_1+1, n-k-r_1+1) \right\}^{-\frac{1}{c}} \tag{4.19}$$

and $$\hat{p}_{JG} = \left\{ \frac{\Gamma(r_1)\Gamma(r_2)}{H_3} \sum_{k=0}^{n-r} \binom{n-r}{k} A_{2k}^{-(r_1)} B_{2k}^{-(r_2)} \text{Beta}(k+r_1-c+1, n-k-r_1+1) \right\}^{-\frac{1}{c}}. \tag{4.20}$$

### 4.4.3. Bayes Estimators with Importance Sampling Procedure

The joint Jeffreys prior of parameters given in (4.14) is incorporated with the likelihood (3.1) to produce the following expression

$$\psi_3(\boldsymbol{\theta}|\mathbf{x}) \propto p^{r_1}(1-p)^{r_2} \theta_1^{r_1-1} \theta_2^{r_2-1} \exp\left(-\theta_1 \sum_{j=1}^{r_1} x_{1j}\right)$$

$$\times \exp\left\{-\theta_2 \sum_{j=1}^{r_2} \ln\left(1+\frac{x_{2j}}{\delta}\right)\right\} \left\{ pe^{-\theta_1 t} + (1-p)\left(1+\frac{t}{\delta}\right)^{-\theta_2} \right\}^{n-r} \tag{4.21}$$

$$\psi_3(\boldsymbol{\theta}|\mathbf{x}) \propto \Gamma_P\left(r_1, \sum_{j=1}^{r_1} x_{1j}\right) \Gamma_P\left(r_2, \sum_{j=1}^{r_2} \ln\left(1+\frac{x_{2j}}{\delta}\right)\right) \text{Beta}_P(r_1+1, r_2+1) h(\boldsymbol{\theta}), \tag{4.22}$$

where $h(\boldsymbol{\theta})$ is defined in (4.10). Following the procedure of determining approximate Bayes estimators described in Section 4.3.3, the expressions of approximate Bayes estimators under SELF are

$$\hat{\theta}_{iISJS} = \frac{\sum_{k=1}^{1000} \theta_{ik} h(\theta_{1k}, \theta_{2k}, p_k)}{\sum_{k=1}^{1000} h(\theta_{1k}, \theta_{2k}, p_k)}, \quad i=1,2 \quad \text{and} \quad \hat{p}_{ISJS} = \frac{\sum_{k=1}^{1000} p_k h(\theta_{1k}, \theta_{2k}, p_k)}{\sum_{k=1}^{1000} h(\theta_{1k}, \theta_{2k}, p_k)}.$$

And the approximate Bayes estimators under GELF are

$$\hat{\theta}_{iISJG} = \left\{ \frac{\sum_{k=1}^{1000} \theta_{ik}^{-c} h(\theta_{1k}, \theta_{2k}, p_k)}{\sum_{k=1}^{1000} h(\theta_{1k}, \theta_{2k}, p_k)} \right\}^{-\frac{1}{c}}, \ i = 1, 2 \quad \text{and} \quad \hat{p}_{ISJG} = \left\{ \frac{\sum_{k=1}^{1000} p_k^{-c} h(\theta_{1k}, \theta_{2k}, p_k)}{\sum_{k=1}^{1000} h(\theta_{1k}, \theta_{2k}, p_k)} \right\}^{-\frac{1}{c}}.$$

### 4.4.4. Posterior Variances Assuming the Jeffreys Priors

From the marginal posterior distributions assuming Jeffreys priors, the expressions for the posterior variances are

$$Var_3(\theta_1 | \mathbf{x}) = \frac{\Gamma(r_1+2)\Gamma(r_2)}{H_3} \sum_{k=0}^{n-r} \binom{n-r}{k} A_{2k}^{-(r_1+2)} B_{2k}^{-r_2} \text{Beta}(k+r_1+1, n-k-r_1+1)$$

$$- \left\{ \frac{\Gamma(r_1+1)\Gamma(r_2)}{H_3} \sum_{k=0}^{n-r} \binom{n-r}{k} A_{2k}^{-(r_1+1)} B_{2k}^{-r_2} \text{Beta}(k+r_1+1, n-k-r_1+1) \right\}^2, \quad (4.23)$$

$$Var_3(\theta_2 | \mathbf{x}) = \frac{\Gamma(r_1)\Gamma(r_2+2)}{H_3} \sum_{k=0}^{n-r} \binom{n-r}{k} A_{2k}^{-r_1} B_{2k}^{-(r_2+2)} \text{Beta}(k+r_1+1, n-k-r_1+1)$$

$$- \left\{ \frac{\Gamma(r_1)\Gamma(r_2+1)}{H_3} \sum_{k=0}^{n-r} \binom{n-r}{k} A_{2k}^{-r_1} B_{2k}^{-(r_2+1)} \text{Beta}(k+r_1+1, n-k-r_1+1) \right\}^2 \quad (4.24)$$

and $\quad Var_3(p | \mathbf{x}) = \frac{\Gamma(r_1)\Gamma(r_2)}{H_3} \sum_{k=0}^{n-r} \binom{n-r}{k} A_{2k}^{-r_1} B_{2k}^{-r_2} \text{Beta}(k+r_1+3, n-k-r_1+1)$

$$- \left\{ \frac{\Gamma(r_1)\Gamma(r_2)}{H_3} \sum_{k=0}^{n-r} \binom{n-r}{k} A_{2k}^{-r_1} B_{2k}^{-r_2} \text{Beta}(k+r_1+2, n-k-r_1+1) \right\}^2. \quad (4.25)$$

## 5. Posterior Predictive Distributions

Statistical prediction deals with estimating the future value(s) of observed random variable using the limited information available at hand. Bayesian statistics provides techniques of predicting the future value(s) of observed random variable after observing only a single random variable. We derive posterior predictive distributions and obtain Bayes point predictors as well as predictive intervals.

### 5.1. Posterior Predictive Distribution Assuming the Uniform Priors

Using the joint posterior distribution assuming the uniform priors and the mixture model (2.4), the predictive distribution for the future observation $Y = X_{n+1}$ given data is defined as

$$f(y | \mathbf{x}) = \int_0^1 \int_0^\infty \int_0^\infty g_1(\mathbf{\theta} | \mathbf{x}) f(y | \mathbf{\theta}) d\theta_1 d\theta_2 dp,$$

After substituting the values and simplification, we get

$$f(y|\underline{x}) = \frac{1}{H_2}\left[\Gamma(r_1+2)\Gamma(r_2+1)\sum_{k=0}^{n-r}\binom{n-r}{k}(A_{2k}+y)^{-(r_1+2)}B_{2k}^{-(r_2+1)}\right.$$

$$\times \text{Beta}(k+r_1+2, n-k-r_1+1) + \frac{\Gamma(r_1+1)\Gamma(r_2+2)}{y+\delta}$$

$$\left.\times \sum_{k=0}^{n-r}\binom{n-r}{k}A_{2k}^{-(r_1+1)}\left\{B_{2k}+\ln\left(1+\frac{y}{\delta}\right)\right\}^{-(r_2+2)}\text{Beta}(k+r_1+1, n-k-r_1+2)\right], \quad (4.26)$$

where $A_{2k}$, $B_{2k}$ and $H_2$ are defined as above. Suppose $L$ and $U$ be the lower and upper bounds of Bayesian predictive interval. A $100(1-\alpha)\%$ predictive interval i.e., $(L,U)$ is obtained as

$$\int_0^L f(y|\mathbf{x})dy = \frac{\alpha}{2} = \int_U^\infty f(y|\mathbf{x})dy.$$

After some simplifications, these equations may also be expressed as

$$\frac{\alpha}{2} = \frac{1}{H_2}\left[\frac{\Gamma(r_1+2)\Gamma(r_2+1)}{r_1+1}\sum_{k=0}^{n-r}\binom{n-r}{k}\left\{A_{2k}^{-(r_1+1)} - (A_{2k}+L)^{-(r_1+1)}\right\}B_{2k}^{-(r_2+1)}\right.$$

$$\times \text{Beta}(k+r_1+2, n-k-r_1+1) + \frac{\Gamma(r_1+1)\Gamma(r_2+2)}{r_2+1}$$

$$\left.\times \sum_{k=0}^{n-r}\binom{n-r}{k}A_{2k}^{-(r_1+1)}\left\{B_{2k}^{-(r_2+1)} - \left(B_{2k}+\ln\left(1+\frac{L}{\delta}\right)\right)^{-(r_2+1)}\right\}\text{Beta}(k+r_1+1, n-k-r_1+2)\right] \quad (4.27)$$

and

$$\frac{\alpha}{2} = \frac{1}{H_2}\left[\frac{\Gamma(r_1+2)\Gamma(r_2+1)}{r_1+1}\sum_{k=0}^{n-r}\binom{n-r}{k}(A_{2k}+U)^{-(r_1+1)}B_{2k}^{-(r_2+1)}\right.$$

$$\times \text{Beta}(k+r_1+2, n-k-r_1+1) + \frac{\Gamma(r_1+1)\Gamma(r_2+2)}{r_2+1}$$

$$\left.\times \sum_{k=0}^{n-r}\binom{n-r}{k}A_{2k}^{-(r_1+1)}\left\{B_{2k}+\ln\left(1+\frac{U}{\delta}\right)\right\}^{-(r_2+1)}\text{Beta}(k+r_1+1, n-k-r_1+2)\right]. \quad (4.28)$$

### 5.2. Bayes Point Predictor Assuming the Uniform Priors

Bayes point predictor (median) is obtained using the posterior predictive distribution (4.26). A solution of the following equation gives Bayes point predictor ($M$)

$$\frac{1}{2} = \frac{1}{H_2}\left[\frac{\Gamma(r_1+2)\Gamma(r_2+1)}{r_1+1}\sum_{k=0}^{n-r}\binom{n-r}{k}\left\{A_{2k}^{-(r_1+1)} - (A_{2k}+M)^{-(r_1+1)}\right\}B_{2k}^{-(r_2+1)}\right.$$

$$\times \text{Beta}(k+r_1+2, n-k-r_1+1) + \frac{\Gamma(r_1+1)\Gamma(r_2+2)}{r_2+1}$$

$$\times \sum_{k=0}^{n-r} \binom{n-r}{k} A_{2k}^{-(r_1+1)} \left\{ B_{2k}^{-(r_2+1)} - \left( B_{2k} + \ln\left(1+\frac{M}{\delta}\right) \right)^{-(r_2+1)} \right\} \text{Beta}(k+r_1+1, n-k-r_1+2) \Bigg]. \quad (4.29)$$

### 5.3. Posterior Predictive Distribution Assuming the Jeffreys Priors

Using the joint posterior distribution assuming the Jeffreys priors and the mixture model (2.4), predictive distribution for the future observation $Y = X_{n+1}$ given data is defined as

$$f(y|\mathbf{x}) = \frac{1}{H_3} \Bigg[ \Gamma(r_1+1)\Gamma(r_2) \sum_{k=0}^{n-r} \binom{n-r}{k} (A_{2k}+y)^{-(r_1+1)} B_{2k}^{-r_2}$$

$$\times \text{Beta}(k+r_1+2, n-k-r_1+1) + \frac{\Gamma(r_1)\Gamma(r_2+1)}{y+\delta}$$

$$\times \sum_{k=0}^{n-r} \binom{n-r}{k} A_{2k}^{-r_1} \left\{ B_{2k} + \ln\left(1+\frac{y}{\delta}\right) \right\}^{-(r_2+1)} \text{Beta}(k+r_1+1, n-k-r_1+2) \Bigg], \quad (4.30)$$

where $A_{2k}$, $B_{2k}$ and $H_3$ are defined as above. A $100(1-\alpha)\%$ predictive interval i.e., $(L,U)$ is obtained as

$$\frac{\alpha}{2} = \frac{1}{H_3} \Bigg[ \frac{\Gamma(r_1+1)\Gamma(r_2)}{r_1} \sum_{k=0}^{n-r} \binom{n-r}{k} \left\{ A_{2k}^{-r_1} - (A_{2k}+L)^{-r_1} \right\} B_{2k}^{-r_2}$$

$$\times \text{Beta}(k+r_1+2, n-k-r_1+1) + \frac{\Gamma(r_1)\Gamma(r_2+1)}{r_2}$$

$$\times \sum_{k=0}^{n-r} \binom{n-r}{k} A_{2k}^{-r_1} \left\{ B_{2k}^{-r_2} - \left( B_{2k} + \ln\left(1+\frac{L}{\delta}\right) \right)^{-r_2} \right\} \text{Beta}(k+r_1+1, n-k-r_1+2) \Bigg] \quad (4.31)$$

and 
$$\frac{\alpha}{2} = \frac{1}{H_3} \Bigg[ \frac{\Gamma(r_1+1)\Gamma(r_2)}{r_1} \sum_{k=0}^{n-r} \binom{n-r}{k} B_{2k}^{-r_2} (A_{2k}+U)^{-r_1}$$

$$\times \text{Beta}(k+r_1+2, n-k-r_1+1) + \frac{\Gamma(r_1)\Gamma(r_2+1)}{r_2}$$

$$\times \sum_{k=0}^{n-r} \binom{n-r}{k} A_{2k}^{-r_1} \left\{ B_{2k} + \ln\left(1+\frac{U}{\delta}\right) \right\}^{-r_2} \text{Beta}(k+r_1+1, n-k-r_1+2) \Bigg]. \quad (4.32)$$

### 5.4. Bayes Point Predictor Assuming the Jeffreys Priors

Bayes point predictor (median) is obtained using the posterior predictive distribution (4.26). A solution of the following equation gives Bayes point predictor ($M$)

$$\frac{1}{2} = \frac{1}{H_3}\left[\frac{\Gamma(r_1+1)\Gamma(r_2)}{r_1}\sum_{k=0}^{n-r}\binom{n-r}{k}\left\{A_{2k}^{-r_1} - (A_{2k}+M)^{-r_1}\right\}B_{2k}^{-r_2}\right.$$

$$\times \text{Beta}(k+r_1+2, n-k-r_1+1) + \frac{\Gamma(r_1)\Gamma(r_2+1)}{r_2}$$

$$\left.\times \sum_{k=0}^{n-r}\binom{n-r}{k}A_{2k}^{-r_1}\left\{B_{2k}^{-r_2} - \left(B_{2k}+\ln\left(1+\frac{M}{\delta}\right)\right)^{-r_2}\right\}\text{Beta}(k+r_1+1, n-k-r_1+2)\right]. \quad (4.33)$$

## 6. A Simulation Study

Following the procedure of simulation from Saleem et al. (2010), samples of different sizes $n = 25, 50, 100$ and $200$ are generated from the mixture model given in (2.4). The censoring times $T \in (0.30, 0.40)$ are chosen for the parameters choice $(\theta_1, \theta_2) \in (10, 10)$ and $p \in (0.40, 0.50)$. Each time 10000 samples are generated to obtain estimates as well as their estimated risks with the help of Mathematica software.

Tables 6.1−6.6 show the estimates and their estimated risks assuming uniform priors. In Tables 6.1 and 6.4, the estimates and their risks are computed under SELF. For small sample size, the ML method provides accurate estimates with small estimated risks as compared to the Bayes estimates. Actual and approximate Bayes estimates computed under SELF are approximately same and have small estimated risks. The estimates and their estimated risks are computed under GELF for $c = 1.2$ and $c = -1.2$ are shown in Tables 6.2 and 6.3, respectively. The performance of Bayes estimates computed under GELF are slightly poor than the Bayes estimates computed under SELF, however, the estimated risks are very small under GELF. It is also observed that the Bayes estimate of $\theta_2$ is over estimated for $c = 1.2$ in most of the cases. As the value of $T$ increases, the estimates get more close to the assumed values and their estimated risks also decrease. The performance of proportion $p$ is good under SELF with small estimated risk. Hence, it is concluded that the ML method gives more accurate estimates for the small sample size as compared to the Bayes estimates assuming uniform priors. For large sample size, ML estimates do not differ much from Bayes estimates. Also, the Bayes estimates are precise under SELF but estimated risks are small under GELF.

Tables 6.7−6.12 show the estimates and their estimated risks assuming Jeffreys priors. In Tables 6.7 and 6.10, the ML method provides accurate estimates for small sample size. But when the sample size increases the Bayes estimates assuming Jeffreys priors give more accurate estimates than ML estimates with small estimated risks. Actual and approximate Bayes estimates do not differ much but approximate Bayes estimates are precise than actual Bayes estimates. Large values of time $T$ result in good agreement with assumed values of the parameters. The approximate Bayes estimates have the smallest estimated risks for large values of $T$. Tables 6.8 and 6.9 show the estimates and their estimated risks computed under GELF for $c = 1.2$ and $c = -1.2$ respectively. For $c = -1.2$, the under estimation is considered more serious but $\theta_1$ is under estimated in some cases using small sample sizes. And the over estimation is considered more serious for $c = 1.2$ but $\theta_2$ is over estimated in some cases using large sample sizes. Approximate Bayes estimates assuming Jeffreys priors are more precise than actual Bayes estimates. Approximate Bayes estimates are more reliable for medium and large sample sizes. The estimated risks of estimates are very small under GELF than under SELF and the risk decreases by increasing $n$ and $T$. It is observed that the approximate Bayes estimates are better in estimating the parameters $\theta_1$ and $\theta_2$ when the Jeffreys priors are assumed.

Table 6.13 shows the Bayes point predictors (medians) and 99% Bayesian predictive intervals. The length of predictive intervals assuming uniform priors are small for medium and large sample sizes. And the length decreases by increasing the value of $T$. The predictive intervals assuming Jeffreys prior are wider than the intervals assuming uniform priors. More the proportion of exponential data in the mixture data, narrower the length of predictive interval becomes.

## 7. A Real Life Example

The analysis of mixture model (2.4) is carried out using a data set taken form Mendenhall and Hader (1958). The data represent the failure times of radio transmitter receivers in a commercial airline. The failure time is set at 630 hours due to general policy of the airline. The first 100 observations of mixture data seem to follow the mixture model developed in this study with parameters $\theta_1 = 0.00476$, $\theta_2 = 149.370$ and $\delta = 28397$. Since $\delta$ is considered a known constant, so our job is to estimate $\theta_1$ and $\theta_2$ only. The parameters are estimated using ML and Bayesian methods along with their estimated risks.

Tables 7.1−7.3 show the estimates and their estimated risks using real data set. The ML method underestimates the parameters with larger risks. For small test time $T$, the ML method fails to estimate the parameters precisely while Bayes estimators have small estimated risks. The risks of Bayes estimators decrease quickly when the termination time increases. Among the estimators, the performance of parameter of Lomax distribution ($\theta_2$) is poor with larger risk. The estimated risks of Bayes estimators assuming Jeffreys priors are relatively large. The estimated risks of ML and Bayes estimators under GELF are significantly small. For $c = 1.2$, there is only one instance where the parameter $\theta_2$ is over estimated. The estimated risks show decreasing trend when $T$ increases. The ML and approximate Bayes estimators have slightly larger risks for $c = -1.2$. And Bayes estimators underestimate the parameters but retain less difference between assumed parameters and the Bayes estimates.

Table 7.4 presents the Bayes point predictor and 99% Bayesian predictive intervals using real data set. For a maximum value of $T = 630$ hours, the median predicted lives of transmitter receivers are 137 hours and 140 hours assuming uniform and Jeffreys priors respectively. And the predicted lives are expected to fall between 0.98 hours to 1116 hours using uniform priors and between 1 hour to 1139 hours using Jeffreys priors. It is observed that the length of predicted interval is wide using Jeffreys priors.

## 8. Conclusion of the Study

In this study, a two-component mixture of two different lifetime distributions is proposed i.e., exponential and Lomax distributions. For small sample size, the ML method provides accurate estimates with small estimated risks. Actual Bayes estimators perform better for medium and large sample sizes and their estimated risks are small. The parameter $\theta_2$ is underestimated in some cases for $c = 1.2$, assuming uniform priors. We recommend actual Bayes estimators when uniform priors are taken and approximate Bayes estimators otherwise. Bayes estimators perform better for large values of $T$. The estimated risks of component parameters $(\theta_1, \theta_2)$ are much small under GELF and the estimated risk of parameter $p$ is small under SELF. For the estimation of component parameters, it is observed that Jeffreys priors are preferable over uniform priors. Another interesting remark is the reduction in risk of component parameter by increasing the proportion of that component in the mixture data.

For real data set, the ML method underestimates the parameters and their estimated risks are large. The estimated risks of Bayes estimators decrease quickly when the termination time increases. Among the estimators, the parameter $\theta_2$ has poor estimate and high estimated risk as well. The estimated risks of Bayes estimators assuming Jeffreys priors are relatively large when computed under SELF. The estimated risks computed under GELF are significantly small. The estimated risks show decreasing trend when the termination time increases. The real data set show that Bayes estimators perform better than ML estimators. Observing the predictive estimates, we can say that a maximum functional transmitter receiver can have an average predicted life of 140 hours. The approximate 99% predicted lower bound is 1 hour and 1116 hours is predicted upper bound.

**Appendix:**

Table 6.1: ML Estimates, Bayes Estimates and their Estimated Risks under SELF Assuming the Uniform Priors with Parameters $\theta_1 = 10$, $\theta_2 = 10$ and $p = 0.4$.

| T | n | ML Estimates | | | Approximate Bayes Estimates | | | Actual Bayes Estimates | | |
|---|---|---|---|---|---|---|---|---|---|---|
| | | $\hat{\theta}_{1ML}$ | $\hat{\theta}_{2ML}$ | $\hat{p}_{ML}$ | $\hat{\theta}_{1ISUS}$ | $\hat{\theta}_{2ISUS}$ | $\hat{p}_{ISUS}$ | $\hat{\theta}_{1US}$ | $\hat{\theta}_{2US}$ | $\hat{p}_{US}$ |
| 0.4 | 25 | 9.94190 | 9.92075 | 0.39973 | 10.83127 | 10.82868 | 0.40962 | 10.85804 | 10.83304 | 0.40940 |
| | | **2.71689** | **2.58668** | **0.00021** | **3.74296** | **3.87495** | **0.00027** | **3.71213** | **3.85748** | **0.00027** |
| | 50 | 9.97203 | 10.02542 | 0.40026 | 10.39271 | 10.48613 | 0.40507 | 10.44143 | 10.48244 | 0.40531 |
| | | **2.46188** | **2.09489** | **0.00014** | **2.60881** | **2.43809** | **0.00017** | **2.61155** | **2.33307** | **0.00014** |
| | 100 | 10.00462 | 10.11969 | 0.40081 | 10.26762 | 10.36976 | 0.40338 | 10.23706 | 10.36261 | 0.40352 |
| | | **2.01693** | **1.63145** | **0.00008** | **2.09513** | **1.73647** | **0.00010** | **2.01581** | **1.73645** | **0.00008** |
| | 200 | 10.07153 | 10.09974 | 0.40038 | 10.16073 | 10.28195 | 0.40199 | 10.18235 | 10.22917 | 0.40186 |
| | | **1.40577** | **1.00183** | **0.00004** | **1.45691** | **1.05285** | **0.00007** | **1.40837** | **1.04144** | **0.00004** |
| 0.5 | 25 | 9.89073 | 9.89064 | 0.39993 | 10.81968 | 10.68577 | 0.40853 | 10.80154 | 10.69371 | 0.40854 |
| | | **2.65203** | **2.48076** | **0.00007** | **3.77185** | **3.67141** | **0.00016** | **3.67959** | **3.50240** | **0.00014** |
| | 50 | 9.97558 | 9.97751 | 0.40019 | 10.40386 | 10.43465 | 0.40464 | 10.43020 | 10.39033 | 0.40480 |
| | | **2.33753** | **2.01100** | **0.00005** | **2.59134** | **2.40947** | **0.00008** | **2.61669** | **2.23528** | **0.00007** |
| | 100 | 10.03285 | 10.05201 | 0.40017 | 10.25803 | 10.28567 | 0.40274 | 10.25639 | 10.26701 | 0.40260 |
| | | **1.86271** | **1.53114** | **0.00003** | **2.04711** | **1.60103** | **0.00004** | **1.94460** | **1.61951** | **0.00003** |
| | 200 | 10.07638 | 10.08245 | 0.40018 | 10.18153 | 10.22054 | 0.40158 | 10.18575 | 10.19348 | 0.40145 |
| | | **1.33166** | **0.90119** | **0.00002** | **1.36556** | **0.95999** | **0.00002** | **1.36528** | **0.93875** | **0.00002** |

Table 6.2: ML Estimates, Bayes Estimates and their Estimated Risks under GELF Assuming the Uniform Priors with Parameters $\theta_1 = 10$, $\theta_2 = 10$, $p = 0.4$ and $c = 1.2$.

| T | n | ML Estimates | | | Approximate Bayes Estimates | | | Actual Bayes Estimates | | |
|---|---|---|---|---|---|---|---|---|---|---|
| | | $\hat{\theta}_{1ML}$ | $\hat{\theta}_{2ML}$ | $\hat{p}_{ML}$ | $\hat{\theta}_{1ISUG}$ | $\hat{\theta}_{2ISUG}$ | $\hat{p}_{ISUG}$ | $\hat{\theta}_{1UG}$ | $\hat{\theta}_{2UG}$ | $\hat{p}_{UG}$ |
| 0.4 | 25 | 9.89959 | 9.94443 | 0.39988 | 9.56690 | 9.99357 | 0.38378 | 9.55737 | 9.99027 | 0.38374 |
| | | **0.01980** | **0.01884** | **0.00093** | **0.02071** | **0.01806** | **0.00221** | **0.02067** | **0.01778** | **0.00214** |
| | 50 | 9.93816 | 10.02108 | 0.40035 | 9.73485 | 10.05304 | 0.39232 | 9.72900 | 10.04334 | 0.39221 |
| | | **0.01824** | **0.01570** | **0.00062** | **0.01771** | **0.01472** | **0.00091** | **0.01743** | **0.01461** | **0.00084** |
| | 100 | 9.99920 | 10.08621 | 0.40076 | 9.88023 | 10.12501 | 0.39687 | 9.87453 | 10.10424 | 0.39675 |
| | | **0.01507** | **0.01173** | **0.00037** | **0.01442** | **0.01117** | **0.00047** | **0.01422** | **0.01117** | **0.00038** |
| | 200 | 10.05374 | 10.10013 | 0.40050 | 10.00579 | 10.15157 | 0.39878 | 9.98170 | 10.11642 | 0.39859 |
| | | **0.01034** | **0.00716** | **0.00020** | **0.01002** | **0.00700** | **0.00031** | **0.00997** | **0.00698** | **0.00019** |
| 0.5 | 25 | 9.90907 | 9.90089 | 0.39969 | 9.63868 | 9.92899 | 0.38300 | 9.63428 | 9.92363 | 0.38303 |
| | | **0.02009** | **0.01866** | **0.00038** | **0.02114** | **0.01858** | **0.00178** | **0.02112** | **0.01842** | **0.00171** |
| | 50 | 9.95736 | 9.99371 | 0.40040 | 9.79373 | 10.00626 | 0.39196 | 9.79239 | 10.00387 | 0.39201 |
| | | **0.01753** | **0.01511** | **0.00022** | **0.01735** | **0.01449** | **0.00055** | **0.01721** | **0.01440** | **0.00050** |
| | 100 | 10.00636 | 10.09180 | 0.40042 | 9.92352 | 10.10723 | 0.39625 | 9.91801 | 10.09949 | 0.39621 |
| | | **0.01472** | **0.01036** | **0.00013** | **0.01444** | **0.01008** | **0.00022** | **0.01437** | **0.01006** | **0.00019** |
| | 200 | 10.06737 | 10.10150 | 0.40024 | 10.01846 | 10.11735 | 0.39828 | 10.01609 | 10.11043 | 0.39819 |
| | | **0.00931** | **0.00630** | **0.00007** | **0.00919** | **0.00620** | **0.00010** | **0.00916** | **0.00623** | **0.00008** |

Table 6.3: ML Estimates, Bayes Estimates and their Estimated Risks under GELF Assuming the Uniform Priors with Parameters $\theta_1 = 10$, $\theta_2 = 10$, $p = 0.4$ and $c = -1.2$.

| T | n | ML Estimates | | | Approximate Bayes Estimates | | | Actual Bayes Estimates | | |
|---|---|---|---|---|---|---|---|---|---|---|
| | | $\hat{\theta}_{1ML}$ | $\hat{\theta}_{2ML}$ | $\hat{p}_{ML}$ | $\hat{\theta}_{1ISUG}$ | $\hat{\theta}_{2ISUG}$ | $\hat{p}_{ISUG}$ | $\hat{\theta}_{1UG}$ | $\hat{\theta}_{2UG}$ | $\hat{p}_{UG}$ |
| 0.4 | 25 | 9.89748 | 9.90553 | 0.39942 | 10.94156 | 10.87540 | 0.41116 | 10.93392 | 10.87566 | 0.41117 |
| | | **0.02143** | **0.02025** | **0.00104** | **0.02138** | **0.02014** | **0.00133** | **0.02125** | **0.01999** | **0.00126** |
| | 50 | 9.98437 | 9.98520 | 0.40018 | 10.51008 | 10.49545 | 0.40654 | 10.50576 | 10.49616 | 0.40658 |
| | | **0.01927** | **0.01651** | **0.00064** | **0.01701** | **0.01534** | **0.00074** | **0.01687** | **0.01526** | **0.00068** |
| | 100 | 10.03181 | 10.07862 | 0.40067 | 10.29665 | 10.35579 | 0.40422 | 10.29469 | 10.34436 | 0.40401 |
| | | **0.01589** | **0.01154** | **0.00038** | **0.01442** | **0.01117** | **0.00047** | **0.01428** | **0.01112** | **0.00039** |
| | 200 | 10.06508 | 10.11790 | 0.40045 | 10.20963 | 10.28186 | 0.40224 | 10.19294 | 10.25689 | 0.40222 |
| | | **0.01064** | **0.00673** | **0.00018** | **0.01019** | **0.00672** | **0.00030** | **0.01008** | **0.00672** | **0.00019** |
| 0.5 | 25 | 9.86741 | 9.89690 | 0.40006 | 10.88592 | 10.76596 | 0.41066 | 10.88822 | 10.76196 | 0.41072 |
| | | **0.02187** | **0.01951** | **0.00031** | **0.02202** | **0.01956** | **0.00077** | **0.02193** | **0.01945** | **0.00074** |
| | 50 | 9.99304 | 10.00538 | 0.40003 | 10.50419 | 10.45535 | 0.40576 | 10.50447 | 10.45288 | 0.40577 |
| | | **0.01862** | **0.01561** | **0.00023** | **0.01744** | **0.01484** | **0.00038** | **0.01737** | **0.01480** | **0.00034** |
| | 100 | 10.03264 | 10.08778 | 0.40019 | 10.28633 | 10.32265 | 0.40316 | 10.28762 | 10.31833 | 0.40317 |
| | | **0.01483** | **0.01057** | **0.00013** | **0.01407** | **0.01040** | **0.00019** | **0.01399** | **0.01038** | **0.00016** |
| | 200 | 10.07407 | 10.08402 | 0.40031 | 10.20106 | 10.20775 | 0.40185 | 10.19798 | 10.20445 | 0.40188 |
| | | **0.00949** | **0.00640** | **0.00007** | **0.00928** | **0.00640** | **0.00010** | **0.00927** | **0.00640** | **0.00008** |

Table 6.4: ML Estimates, Bayes Estimates and their Estimated Risks under SELF Assuming the Uniform Priors with Parameters $\theta_1 = 10$, $\theta_2 = 10$ and $p = 0.5$.

| T | n | ML Estimates | | | Approximate Bayes Estimates | | | Actual Bayes Estimates | | |
|---|---|---|---|---|---|---|---|---|---|---|
| | | $\hat{\theta}_{1ML}$ | $\hat{\theta}_{2ML}$ | $\hat{p}_{ML}$ | $\hat{\theta}_{1ISUS}$ | $\hat{\theta}_{2ISUS}$ | $\hat{p}_{ISUS}$ | $\hat{\theta}_{1US}$ | $\hat{\theta}_{2US}$ | $\hat{p}_{US}$ |
| 0.4 | 25 | 9.74755 | 9.04876 | 0.49151 | 10.33249 | 10.25324 | 0.50669 | 10.30824 | 10.61287 | 0.49812 |
| | | **2.89694** | **3.21290** | **0.00029** | **2.87017** | **1.48294** | **0.00017** | **2.73585** | **5.49818** | **0.00018** |
| | 50 | 10.00643 | 9.98718 | 0.49995 | 10.55837 | 11.39211 | 0.50362 | 10.39023 | 10.52444 | 0.50092 |
| | | **2.30801** | **2.33901** | **0.00015** | **2.83954** | **3.63147** | **0.00007** | **2.40892** | **2.65797** | **0.00013** |
| | 100 | 10.05847 | 10.07153 | 0.50021 | 9.19212 | 10.53773 | 0.49663 | 10.24846 | 10.35642 | 0.50083 |
| | | **1.84342** | **1.82859** | **0.00009** | **1.16588** | **2.64760** | **0.00003** | **1.84158** | **1.92440** | **0.00008** |
| | 200 | 10.07012 | 10.11251 | 0.50027 | 11.10574 | 10.77488 | 0.50513 | 10.16460 | 10.26173 | 0.50063 |
| | | **1.18208** | **1.14247** | **0.00005** | **2.33408** | **0.91508** | **0.00005** | **1.17875** | **1.18704** | **0.00004** |
| 0.5 | 25 | 9.69237 | 8.73879 | 0.48959 | 10.04020 | 10.40688 | 0.49728 | 10.07667 | 10.46593 | 0.49727 |
| | | **3.00425** | **3.53382** | **0.00022** | **2.72693** | **5.58610** | **0.00015** | **2.70799** | **5.72162** | **0.00012** |
| | 50 | 9.98550 | 9.97948 | 0.50024 | 10.35955 | 10.48101 | 0.50064 | 10.35345 | 10.46244 | 0.50084 |
| | | **2.25528** | **2.23378** | **0.00006** | **2.39096** | **2.60643** | **0.00006** | **2.42638** | **2.51531** | **0.00005** |
| | 100 | 10.08498 | 10.07819 | 0.50012 | 10.23572 | 10.29186 | 0.50048 | 10.26711 | 10.33318 | 0.50051 |
| | | **1.75388** | **1.71472** | **0.00003** | **1.77583** | **1.77861** | **0.00004** | **1.81092** | **1.83481** | **0.00003** |
| | 200 | 10.06994 | 10.08069 | 0.50016 | 10.14757 | 10.22963 | 0.50057 | 10.15957 | 10.21109 | 0.50038 |
| | | **1.10177** | **1.08271** | **0.00002** | **1.14013** | **1.16086** | **0.00002** | **1.12054** | **1.12448** | **0.00002** |

Table 6.5: ML Estimates, Bayes Estimates and their Estimated Risks under GELF Assuming the Uniform Priors with Parameters $\theta_1 = 10$, $\theta_2 = 10$, $p = 0.5$ and $c = 1.2$.

| T | n | ML Estimates | | | Approximate Bayes Estimates | | | Actual Bayes Estimates | | |
|---|---|---|---|---|---|---|---|---|---|---|
| | | $\hat{\theta}_{1ML}$ | $\hat{\theta}_{2ML}$ | $\hat{p}_{ML}$ | $\hat{\theta}_{1ISUG}$ | $\hat{\theta}_{2ISUG}$ | $\hat{p}_{ISUG}$ | $\hat{\theta}_{1UG}$ | $\hat{\theta}_{2UG}$ | $\hat{p}_{UG}$ |
| 0.4 | 25 | 9.66722 | 9.05286 | 0.49152 | 9.06313 | 9.47625 | 0.47592 | 9.04546 | 9.43064 | 0.47569 |
| | | **0.02235** | **0.02616** | **0.00088** | **0.02561** | **0.03079** | **0.00245** | **0.02539** | **0.02832** | **0.00234** |
| | 50 | 9.99429 | 9.92220 | 0.50003 | 9.84213 | 9.95345 | 0.49019 | 9.83457 | 9.93681 | 0.49006 |
| | | **0.01735** | **0.01718** | **0.00043** | **0.01637** | **0.01603** | **0.00074** | **0.01628** | **0.01584** | **0.00067** |
| | 100 | 10.10064 | 10.03842 | 0.50007 | 10.02117 | 10.06657 | 0.49526 | 10.01313 | 10.04662 | 0.49506 |
| | | **0.01314** | **0.01305** | **0.00026** | **0.01240** | **0.01237** | **0.00034** | **0.01229** | **0.01229** | **0.00030** |
| | 200 | 10.08019 | 10.09835 | 0.50026 | 10.04918 | 10.14502 | 0.49808 | 10.03134 | 10.10805 | 0.49781 |
| | | **0.00878** | **0.00844** | **0.00014** | **0.00839** | **0.00816** | **0.00019** | **0.00840** | **0.00813** | **0.00015** |
| 0.5 | 25 | 9.73659 | 8.70397 | 0.48924 | 8.93768 | 9.27485 | 0.47491 | 8.92382 | 9.24127 | 0.47486 |
| | | **0.02359** | **0.02962** | **0.00068** | **0.02962** | **0.03147** | **0.00239** | **0.02955** | **0.02990** | **0.00225** |
| | 50 | 9.98490 | 9.97717 | 0.50021 | 9.86928 | 9.98336 | 0.48990 | 9.86493 | 9.97425 | 0.48993 |
| | | **0.01643** | **0.01616** | **0.00017** | **0.01586** | **0.01531** | **0.00048** | **0.01585** | **0.01520** | **0.00045** |
| | 100 | 10.05526 | 10.03317 | 0.50007 | 9.99024 | 10.04581 | 0.49500 | 9.98642 | 10.03863 | 0.49496 |
| | | **0.01229** | **0.01241** | **0.00009** | **0.01195** | **0.01194** | **0.00018** | **0.01193** | **0.01197** | **0.00016** |
| | 200 | 10.05939 | 10.12158 | 0.50028 | 10.02716 | 10.13167 | 0.49776 | 10.02296 | 10.12723 | 0.49773 |
| | | **0.00819** | **0.00793** | **0.00005** | **0.00806** | **0.00783** | **0.00007** | **0.00804** | **0.00778** | **0.00006** |

Table 6.6: ML Estimates, Bayes Estimates and their Estimated Risks under GELF Assuming the Uniform Priors with Parameters $\theta_1 = 10$, $\theta_2 = 10$, $p = 0.5$ and $c = -1.2$.

| T | n | ML Estimates | | | Approximate Bayes Estimates | | | Actual Bayes Estimates | | |
|---|---|---|---|---|---|---|---|---|---|---|
| | | $\hat{\theta}_{1ML}$ | $\hat{\theta}_{2ML}$ | $\hat{p}_{ML}$ | $\hat{\theta}_{1ISUG}$ | $\hat{\theta}_{2ISUG}$ | $\hat{p}_{ISUG}$ | $\hat{\theta}_{1UG}$ | $\hat{\theta}_{2UG}$ | $\hat{p}_{UG}$ |
| 0.4 | 25 | 9.68755 | 9.02880 | 0.49168 | 10.36385 | 10.74198 | 0.50031 | 10.35312 | 10.75053 | 0.50035 |
| | | **0.02449** | **0.03095** | **0.00088** | **0.01944** | **0.02465** | **0.00064** | **0.01879** | **0.02443** | **0.00051** |
| | 50 | 10.01628 | 9.96110 | 0.50014 | 10.44830 | 10.56084 | 0.50218 | 10.44335 | 10.55547 | 0.50216 |
| | | **0.01793** | **0.01755** | **0.00044** | **0.01609** | **0.01615** | **0.00042** | **0.01592** | **0.01599** | **0.00036** |
| | 100 | 10.07479 | 10.09460 | 0.50030 | 10.29331 | 10.41584 | 0.50151 | 10.29053 | 10.40440 | 0.50140 |
| | | **0.01385** | **0.01358** | **0.00027** | **0.01297** | **0.01292** | **0.00028** | **0.01271** | **0.01291** | **0.00023** |
| | 200 | 10.09005 | 10.08787 | 0.50014 | 10.20729 | 10.27358 | 0.50087 | 10.19708 | 10.24912 | 0.50076 |
| | | **0.00886** | **0.00837** | **0.00014** | **0.00854** | **0.00815** | **0.00018** | **0.00848** | **0.00814** | **0.00013** |
| 0.5 | 25 | 9.71313 | 8.71735 | 0.48936 | 10.18456 | 10.61654 | 0.49916 | 10.18170 | 10.61217 | 0.49925 |
| | | **0.02594** | **0.03532** | **0.00069** | **0.02134** | **0.02576** | **0.00041** | **0.02096** | **0.02548** | **0.00033** |
| | 50 | 9.95977 | 9.93835 | 0.50012 | 10.37057 | 10.46882 | 0.50161 | 10.37002 | 10.46843 | 0.50171 |
| | | **0.01795** | **0.01720** | **0.00018** | **0.01631** | **0.01586** | **0.00018** | **0.01626** | **0.01579** | **0.00015** |
| | 100 | 10.06259 | 10.07506 | 0.50027 | 10.26955 | 10.35087 | 0.50115 | 10.26929 | 10.34949 | 0.50112 |
| | | **0.01259** | **0.01230** | **0.00009** | **0.01203** | **0.01196** | **0.00010** | **0.01199** | **0.01199** | **0.00008** |
| | 200 | 10.05363 | 10.10845 | 0.50016 | 10.15421 | 10.25700 | 0.50066 | 10.15470 | 10.25102 | 0.50064 |
| | | **0.00779** | **0.00772** | **0.00005** | **0.00760** | **0.00771** | **0.00006** | **0.00758** | **0.00769** | **0.00005** |

Table 6.7: ML Estimates, Bayes Estimates and their Estimated Risks under SELF Assuming the Jeffreys Priors with Parameters $\theta_1 = 10$, $\theta_2 = 10$ and $p = 0.4$.

| T | n | ML Estimates | | | Approximate Bayes Estimates | | | Actual Bayes Estimates | | |
|---|---|---|---|---|---|---|---|---|---|---|
| | | $\hat{\theta}_{1ML}$ | $\hat{\theta}_{2ML}$ | $\hat{p}_{ML}$ | $\hat{\theta}_{1ISJS}$ | $\hat{\theta}_{2ISJS}$ | $\hat{p}_{ISJS}$ | $\hat{\theta}_{1JS}$ | $\hat{\theta}_{2JS}$ | $\hat{p}_{JS}$ |
| 0.4 | 25 | 9.94190 | 9.92075 | 0.39973 | 9.79394 | 10.12084 | 0.41026 | 9.78725 | 10.16030 | 0.40991 |
| | | **2.71689** | **2.58668** | **0.00021** | **2.49244** | **2.77349** | **0.00030** | **2.52210** | **2.64754** | **0.00028** |
| | 50 | 9.97203 | 10.02542 | 0.40026 | 9.90624 | 10.16733 | 0.40606 | 9.93353 | 10.13489 | 0.40578 |
| | | **2.46188** | **2.09489** | **0.00014** | **2.21307** | **2.11436** | **0.00017** | **2.17325** | **1.97297** | **0.00015** |
| | 100 | 10.00462 | 10.11969 | 0.40081 | 9.96289 | 10.18121 | 0.40379 | 9.97304 | 10.19606 | 0.40369 |
| | | **2.01693** | **1.63145** | **0.00008** | **1.88944** | **1.56831** | **0.00010** | **1.92080** | **1.58798** | **0.00008** |
| | 200 | 10.07153 | 10.09974 | 0.40038 | 10.05974 | 10.16620 | 0.40204 | 10.03097 | 10.15224 | 0.40210 |
| | | **1.40577** | **1.00183** | **0.00004** | **1.35456** | **0.96436** | **0.00007** | **1.40870** | **0.98754** | **0.00005** |
| 0.5 | 25 | 9.89073 | 9.89064 | 0.39993 | 9.81401 | 10.02936 | 0.40863 | 9.80676 | 10.05540 | 0.40892 |
| | | **2.65203** | **2.48076** | **0.00007** | **2.60340** | **2.70711** | **0.00016** | **2.59733** | **2.59436** | **0.00015** |
| | 50 | 9.97558 | 9.97751 | 0.40019 | 9.87637 | 10.09261 | 0.40526 | 9.88980 | 10.12829 | 0.40511 |
| | | **2.33753** | **2.01100** | **0.00005** | **2.28461** | **2.08761** | **0.00008** | **2.19818** | **2.04915** | **0.00007** |
| | 100 | 10.03285 | 10.05201 | 0.40017 | 10.00608 | 10.17936 | 0.40288 | 10.00244 | 10.13862 | 0.40267 |
| | | **1.86271** | **1.53114** | **0.00003** | **1.92027** | **1.46743** | **0.00004** | **1.83514** | **1.48840** | **0.00003** |
| | 200 | 10.07638 | 10.08245 | 0.40018 | 10.03396 | 10.15592 | 0.40161 | 10.02727 | 10.11983 | 0.40153 |
| | | **1.33166** | **0.90119** | **0.00002** | **1.33317** | **0.95944** | **0.00002** | **1.28362** | **0.91329** | **0.00002** |

Table 6.8: ML Estimates, Bayes Estimates and their Estimated Risks under GELF Assuming the Jeffreys Priors with Parameters $\theta_1 = 10$, $\theta_2 = 10$, $p = 0.4$ and $c = 1.2$.

| T | n | ML Estimates | | | Approximate Bayes Estimates | | | Actual Bayes Estimates | | |
|---|---|---|---|---|---|---|---|---|---|---|
| | | $\hat{\theta}_{1ML}$ | $\hat{\theta}_{2ML}$ | $\hat{p}_{ML}$ | $\hat{\theta}_{1ISJG}$ | $\hat{\theta}_{2ISJG}$ | $\hat{p}_{ISJG}$ | $\hat{\theta}_{1JG}$ | $\hat{\theta}_{2JG}$ | $\hat{p}_{JG}$ |
| 0.4 | 25 | 9.89959 | 9.94443 | 0.39988 | 8.55208 | 9.32108 | 0.38455 | 8.54192 | 9.27862 | 0.38448 |
| | | **0.01980** | **0.01884** | **0.00093** | **0.03664** | **0.02153** | **0.00210** | **0.03693** | **0.02143** | **0.00201** |
| | 50 | 9.93816 | 10.02108 | 0.40035 | 9.21678 | 9.72932 | 0.39271 | 9.20875 | 9.73689 | 0.39263 |
| | | **0.01824** | **0.01570** | **0.00062** | **0.02154** | **0.01521** | **0.00085** | **0.02127** | **0.01514** | **0.00078** |
| | 100 | 9.99920 | 10.08621 | 0.40076 | 9.62187 | 9.97377 | 0.39703 | 9.61635 | 9.95334 | 0.39681 |
| | | **0.01507** | **0.01173** | **0.00037** | **0.01504** | **0.01088** | **0.00045** | **0.01494** | **0.01095** | **0.00037** |
| | 200 | 10.05374 | 10.10013 | 0.40050 | 9.83340 | 10.06756 | 0.39908 | 9.89711 | 10.02748 | 0.39845 |
| | | **0.01034** | **0.00716** | **0.00020** | **0.01046** | **0.00678** | **0.00031** | **0.01026** | **0.00677** | **0.00020** |
| 0.5 | 25 | 9.90907 | 9.90089 | 0.39969 | 8.61515 | 9.24796 | 0.38337 | 8.62827 | 9.24538 | 0.38307 |
| | | **0.02009** | **0.01866** | **0.00038** | **0.03534** | **0.02210** | **0.00169** | **0.03490** | **0.02204** | **0.00170** |
| | 50 | 9.95736 | 9.99371 | 0.40040 | 9.30279 | 9.67282 | 0.39183 | 9.30538 | 9.64838 | 0.39198 |
| | | **0.01753** | **0.01511** | **0.00022** | **0.02014** | **0.01559** | **0.00058** | **0.02068** | **0.01514** | **0.00051** |
| | 100 | 10.00636 | 10.09180 | 0.40042 | 9.62400 | 9.92485 | 0.39634 | 9.66667 | 9.96137 | 0.39634 |
| | | **0.01472** | **0.01036** | **0.00013** | **0.01503** | **0.01030** | **0.00022** | **0.01444** | **0.01029** | **0.00018** |
| | 200 | 10.06737 | 10.10150 | 0.40024 | 9.86312 | 10.02691 | 0.39823 | 9.88461 | 10.00798 | 0.39828 |
| | | **0.00931** | **0.00630** | **0.00007** | **0.00944** | **0.00614** | **0.00010** | **0.00923** | **0.00639** | **0.00008** |

Table 6.9: ML Estimates, Bayes Estimates and their Estimated Risks under GELF Assuming the Jeffreys Priors with Parameters $\theta_1 = 10$, $\theta_2 = 10$, $p = 0.4$ and $c = -1.2$.

| T | n | ML Estimates | | | Approximate Bayes Estimates | | | Actual Bayes Estimates | | |
|---|---|---|---|---|---|---|---|---|---|---|
| | | $\hat{\theta}_{1ML}$ | $\hat{\theta}_{2ML}$ | $\hat{p}_{ML}$ | $\hat{\theta}_{1ISJG}$ | $\hat{\theta}_{2ISJG}$ | $\hat{p}_{ISJG}$ | $\hat{\theta}_{1JG}$ | $\hat{\theta}_{2JG}$ | $\hat{p}_{JG}$ |
| 0.4 | 25 | 9.89748 | 9.90553 | 0.39942 | 9.93542 | 10.24441 | 0.41230 | 9.89065 | 10.23150 | 0.41175 |
| | | **0.02143** | **0.02025** | **0.00104** | **0.01971** | **0.01816** | **0.00141** | **0.02039** | **0.01799** | **0.00134** |
| | 50 | 9.98437 | 9.98520 | 0.40018 | 9.98030 | 10.19941 | 0.40719 | 9.96620 | 10.18476 | 0.40708 |
| | | **0.01927** | **0.01651** | **0.00064** | **0.01750** | **0.01488** | **0.00079** | **0.01710** | **0.01427** | **0.00069** |
| | 100 | 10.03181 | 10.07862 | 0.40067 | 10.02350 | 10.19386 | 0.40397 | 10.01503 | 10.22666 | 0.40406 |
| | | **0.01589** | **0.01154** | **0.00038** | **0.01460** | **0.01083** | **0.00044** | **0.01402** | **0.01068** | **0.00038** |
| | 200 | 10.06508 | 10.11790 | 0.40045 | 10.09684 | 10.19107 | 0.40241 | 10.06013 | 10.14365 | 0.40226 |
| | | **0.01064** | **0.00673** | **0.00018** | **0.01054** | **0.00672** | **0.00030** | **0.01006** | **0.00675** | **0.00021** |
| 0.5 | 25 | 9.86741 | 9.89690 | 0.40006 | 9.87277 | 10.10213 | 0.41092 | 9.91781 | 10.10010 | 0.41094 |
| | | **0.02187** | **0.01951** | **0.00031** | **0.02131** | **0.01794** | **0.00084** | **0.02125** | **0.01909** | **0.00082** |
| | 50 | 9.99304 | 10.00538 | 0.40003 | 9.93375 | 10.10599 | 0.40615 | 9.93718 | 10.09849 | 0.40624 |
| | | **0.01862** | **0.01561** | **0.00023** | **0.01770** | **0.01462** | **0.00038** | **0.01726** | **0.01460** | **0.00036** |
| | 100 | 10.03264 | 10.08778 | 0.40019 | 10.02563 | 10.12650 | 0.40348 | 10.03216 | 10.14354 | 0.40340 |
| | | **0.01483** | **0.01057** | **0.00013** | **0.01392** | **0.01051** | **0.00019** | **0.01406** | **0.01062** | **0.00017** |
| | 200 | 10.07407 | 10.08402 | 0.40031 | 10.06104 | 10.13221 | 0.40189 | 10.08326 | 10.09994 | 0.40177 |
| | | **0.00949** | **0.00640** | **0.00007** | **0.00926** | **0.00629** | **0.00010** | **0.00985** | **0.00633** | **0.00007** |

Table 6.10: ML Estimates, Bayes Estimates and their Estimated Risks under SELF Assuming the Jeffreys Priors with Parameters $\theta_1 = 10$, $\theta_2 = 10$ and $p = 0.5$.

| T | n | ML Estimates | | | Approximate Bayes Estimates | | | Actual Bayes Estimates | | |
|---|---|---|---|---|---|---|---|---|---|---|
| | | $\hat{\theta}_{1ML}$ | $\hat{\theta}_{2ML}$ | $\hat{p}_{ML}$ | $\hat{\theta}_{1ISJS}$ | $\hat{\theta}_{2ISJS}$ | $\hat{p}_{ISJS}$ | $\hat{\theta}_{1JS}$ | $\hat{\theta}_{2JS}$ | $\hat{p}_{JS}$ |
| 0.4 | 25 | 9.74755 | 9.04876 | 0.49151 | 9.45202 | 9.85789 | 0.49844 | 9.44383 | 9.79710 | 0.49832 |
| | | **2.89694** | **3.21290** | **0.00029** | **2.49825** | **4.40878** | **0.00019** | **2.47720** | **4.10812** | **0.00015** |
| | 50 | 10.00643 | 9.98718 | 0.49995 | 9.98339 | 10.12774 | 0.50121 | 10.01174 | 10.09036 | 0.50100 |
| | | **2.30801** | **2.33901** | **0.00015** | **2.13435** | **2.14570** | **0.00014** | **2.09098** | **2.10356** | **0.00012** |
| | 100 | 10.05847 | 10.07153 | 0.50021 | 10.04474 | 10.15863 | 0.50085 | 10.02948 | 10.10557 | 0.50085 |
| | | **1.84342** | **1.82859** | **0.00009** | **1.68836** | **1.71977** | **0.00009** | **1.71502** | **1.69263** | **0.00008** |
| | 200 | 10.07012 | 10.11251 | 0.50027 | 10.06602 | 10.17557 | 0.50092 | 10.06031 | 10.14078 | 0.50050 |
| | | **1.18208** | **1.14247** | **0.00005** | **1.15145** | **1.15262** | **0.00007** | **1.14404** | **1.17480** | **0.00005** |
| 0.5 | 25 | 9.69237 | 8.73879 | 0.48959 | 9.25032 | 9.66399 | 0.49770 | 9.29842 | 9.62103 | 0.49736 |
| | | **3.00425** | **3.53382** | **0.00022** | **2.80012** | **4.41534** | **0.00015** | **2.78060** | **4.37184** | **0.00011** |
| | 50 | 9.98550 | 9.97948 | 0.50024 | 9.97573 | 10.08857 | 0.50085 | 9.92663 | 10.07155 | 0.50104 |
| | | **2.25528** | **2.23378** | **0.00006** | **2.13911** | **2.15057** | **0.00006** | **2.09815** | **2.14013** | **0.00005** |
| | 100 | 10.08498 | 10.07819 | 0.50012 | 10.03071 | 10.13204 | 0.50068 | 10.02744 | 10.10104 | 0.50064 |
| | | **1.75388** | **1.71472** | **0.00003** | **1.68429** | **1.57778** | **0.00004** | **1.69196** | **1.67402** | **0.00003** |
| | 200 | 10.06994 | 10.08069 | 0.50016 | 10.05055 | 10.14487 | 0.50038 | 10.07780 | 10.13414 | 0.50024 |
| | | **1.10177** | **1.08271** | **0.00002** | **1.12281** | **1.05005** | **0.00002** | **1.09747** | **1.06709** | **0.00002** |

Table 6.11: ML Estimates, Bayes Estimates and their Estimated Risks under GELF Assuming the Jeffreys Priors with Parameters $\theta_1 = 10$, $\theta_2 = 10$, $p = 0.5$ and $c = 1.2$.

| T | n | ML Estimates | | | Approximate Bayes Estimates | | | Actual Bayes Estimates | | |
|---|---|---|---|---|---|---|---|---|---|---|
| | | $\hat{\theta}_{1ML}$ | $\hat{\theta}_{2ML}$ | $\hat{p}_{ML}$ | $\hat{\theta}_{1ISJG}$ | $\hat{\theta}_{2ISJG}$ | $\hat{p}_{ISJG}$ | $\hat{\theta}_{1JG}$ | $\hat{\theta}_{2JG}$ | $\hat{p}_{JG}$ |
| 0.4 | 25 | 9.66722 | 9.05286 | 0.49152 | 8.27173 | 8.62589 | 0.47596 | 8.29947 | 8.64193 | 0.47619 |
| | | **0.02235** | **0.02616** | **0.00088** | **0.04136** | **0.03790** | **0.00233** | **0.03998** | **0.03608** | **0.00216** |
| | 50 | 9.99429 | 9.92220 | 0.50003 | 9.48462 | 9.55482 | 0.48987 | 9.43701 | 9.55336 | 0.49020 |
| | | **0.01735** | **0.01718** | **0.00043** | **0.01732** | **0.01732** | **0.00075** | **0.01733** | **0.01664** | **0.00069** |
| | 100 | 10.10064 | 10.03842 | 0.50007 | 9.80892 | 9.87934 | 0.49520 | 9.76246 | 9.83596 | 0.49489 |
| | | **0.01314** | **0.01305** | **0.00026** | **0.01231** | **0.01169** | **0.00036** | **0.01263** | **0.01180** | **0.00031** |
| | 200 | 10.08019 | 10.09835 | 0.50026 | 9.95832 | 10.02677 | 0.49794 | 9.93937 | 10.00433 | 0.49784 |
| | | **0.00878** | **0.00844** | **0.00014** | **0.00799** | **0.00786** | **0.00020** | **0.00855** | **0.00810** | **0.00014** |
| 0.5 | 25 | 9.73659 | 8.70397 | 0.48924 | 8.16730 | 8.44287 | 0.47525 | 8.09921 | 8.48302 | 0.47522 |
| | | **0.02359** | **0.02962** | **0.00068** | **0.04671** | **0.04254** | **0.00226** | **0.04824** | **0.04121** | **0.00216** |
| | 50 | 9.98490 | 9.97717 | 0.50021 | 9.51302 | 9.51514 | 0.48977 | 9.49832 | 9.60892 | 0.48969 |
| | | **0.01643** | **0.01616** | **0.00017** | **0.01716** | **0.01676** | **0.00049** | **0.01735** | **0.01628** | **0.00045** |
| | 100 | 10.05526 | 10.03317 | 0.50007 | 9.81091 | 9.89437 | 0.49511 | 9.78769 | 9.86069 | 0.49514 |
| | | **0.01229** | **0.01241** | **0.00009** | **0.01234** | **0.01177** | **0.00018** | **0.01265** | **0.01195** | **0.00015** |
| | 200 | 10.05939 | 10.12158 | 0.50028 | 9.91120 | 10.00814 | 0.49783 | 9.98069 | 9.98580 | 0.49759 |
| | | **0.00819** | **0.00793** | **0.00005** | **0.00793** | **0.00731** | **0.00008** | **0.00781** | **0.00721** | **0.00006** |

Table 6.12: ML Estimates, Bayes Estimates and their Estimated Risks under GELF Assuming the Jeffreys Priors with Parameters $\theta_1 = 10$, $\theta_2 = 10$, $p = 0.5$ and $c = -1.2$.

| T | n | ML Estimates | | | Approximate Bayes Estimates | | | Actual Bayes Estimates | | |
|---|---|---|---|---|---|---|---|---|---|---|
| | | $\hat{\theta}_{1ML}$ | $\hat{\theta}_{2ML}$ | $\hat{p}_{ML}$ | $\hat{\theta}_{1ISJG}$ | $\hat{\theta}_{2ISJG}$ | $\hat{p}_{ISJG}$ | $\hat{\theta}_{1JG}$ | $\hat{\theta}_{2JG}$ | $\hat{p}_{JG}$ |
| 0.4 | 25 | 9.68755 | 9.02880 | 0.49168 | 9.52513 | 9.97786 | 0.50046 | 9.52824 | 9.96392 | 0.50057 |
| | | **0.02449** | **0.03095** | **0.00088** | **0.02277** | **0.02501** | **0.00053** | **0.02169** | **0.02383** | **0.00046** |
| | 50 | 10.01628 | 9.96110 | 0.50014 | 10.04822 | 10.12267 | 0.50237 | 10.04162 | 10.18201 | 0.50212 |
| | | **0.01793** | **0.01755** | **0.00044** | **0.01647** | **0.01574** | **0.00042** | **0.01578** | **0.01483** | **0.00037** |
| | 100 | 10.07479 | 10.09460 | 0.50030 | 10.07668 | 10.19620 | 0.50148 | 10.05767 | 10.14682 | 0.50130 |
| | | **0.01385** | **0.01358** | **0.00027** | **0.01262** | **0.01232** | **0.00027** | **0.01276** | **0.01236** | **0.00024** |
| | 200 | 10.09005 | 10.08787 | 0.50014 | 10.09722 | 10.16796 | 0.50117 | 10.07301 | 10.13936 | 0.50087 |
| | | **0.00886** | **0.00837** | **0.00014** | **0.00823** | **0.00802** | **0.00019** | **0.00851** | **0.00802** | **0.00013** |
| 0.5 | 25 | 9.71313 | 8.71735 | 0.48936 | 9.38565 | 9.73516 | 0.49881 | 9.37905 | 9.79323 | 0.49941 |
| | | **0.02594** | **0.03532** | **0.00069** | **0.02608** | **0.02540** | **0.00035** | **0.02630** | **0.02664** | **0.00028** |
| | 50 | 9.95977 | 9.93835 | 0.50012 | 10.00717 | 10.13039 | 0.50181 | 10.02326 | 10.10638 | 0.50162 |
| | | **0.01795** | **0.01720** | **0.00018** | **0.01610** | **0.01566** | **0.00017** | **0.01632** | **0.01560** | **0.00015** |
| | 100 | 10.06259 | 10.07506 | 0.50027 | 10.09757 | 10.17474 | 0.50111 | 10.04337 | 10.14392 | 0.50106 |
| | | **0.01259** | **0.01230** | **0.00009** | **0.01231** | **0.01204** | **0.00010** | **0.01266** | **0.01185** | **0.00009** |
| | 200 | 10.05363 | 10.10845 | 0.50016 | 10.08902 | 10.11843 | 0.50069 | 10.07008 | 10.13418 | 0.50067 |
| | | **0.00779** | **0.00772** | **0.00005** | **0.00792** | **0.00739** | **0.00006** | **0.00768** | **0.00750** | **0.00005** |

Table 7: Bayes Point Predictor and 99% Predictive Intervals using Real Life Example.

| | Uniform Priors | | | Jeffreys Priors | | |
|---|---|---|---|---|---|---|
| T | Median | L | U | Median | L | U |
| 300 | 151.908 | 1.06877 | 1358.37 | 156.095 | 1.09894 | 1392.77 |
| 400 | 148.562 | 1.05680 | 1260.57 | 152.106 | 1.08198 | 1291.18 |
| 630 | 137.993 | 0.98939 | 1116.02 | 140.776 | 1.00917 | 1139.74 |

Table 6.13: Bayes Point Predictor and 99% Predictive Intervals for $\theta_1 = 10$ and $\theta_2 = 10$.

| T | n | p = 0.4 | | | p = 0.5 | | |
|---|---|---|---|---|---|---|---|
| | | Median | L | U | Median | L | U |
| | | | | Uniform Priors | | | |
| 0.4 | 25 | 0.00044 | 0.06492 | 0.80122 | 0.00049 | 0.07348 | 1.01608 |
| | 50 | 0.00047 | 0.06790 | 0.72639 | 0.00047 | 0.06769 | 0.71371 |
| | 100 | 0.00048 | 0.06923 | 0.68013 | 0.00048 | 0.06899 | 0.66768 |
| | 200 | 0.00049 | 0.06986 | 0.65871 | 0.00049 | 0.06967 | 0.64444 |
| 0.5 | 25 | 0.00044 | 0.06556 | 0.79377 | 0.00050 | 0.07574 | 1.05148 |
| | 50 | 0.00047 | 0.06781 | 0.71412 | 0.00047 | 0.06755 | 0.69852 |
| | 100 | 0.00049 | 0.06923 | 0.67509 | 0.00049 | 0.06929 | 0.66419 |
| | 200 | 0.00049 | 0.07007 | 0.65737 | 0.00049 | 0.06989 | 0.64339 |
| | | | | Jeffreys Priors | | | |
| 0.4 | 25 | 0.00048 | 0.07079 | 0.89796 | 0.00054 | 0.08090 | 1.17153 |
| | 50 | 0.00049 | 0.07069 | 0.76007 | 0.00049 | 0.07041 | 0.74918 |
| | 100 | 0.00049 | 0.07078 | 0.69821 | 0.00049 | 0.07033 | 0.68307 |
| | 200 | 0.00050 | 0.07082 | 0.66844 | 0.00050 | 0.07052 | 0.65362 |
| 0.5 | 25 | 0.00048 | 0.07108 | 0.88753 | 0.00055 | 0.08249 | 1.19750 |
| | 50 | 0.00049 | 0.07084 | 0.75404 | 0.00049 | 0.07069 | 0.74260 |
| | 100 | 0.00050 | 0.07085 | 0.69328 | 0.00050 | 0.07045 | 0.67783 |
| | 200 | 0.00050 | 0.07069 | 0.66375 | 0.00050 | 0.07046 | 0.64973 |

Table 7.1: ML Estimates, Bayes Estimates and their Estimated Risks under SELF using Real Life Example.

| T | ML Estimates | | | Approximate Bayes Estimates | | | Actual Bayes Estimates | | |
|---|---|---|---|---|---|---|---|---|---|
| | $\hat{\theta}_{1ML}$ | $\hat{\theta}_{2ML}$ | $\hat{p}_{ML}$ | $\hat{\theta}_{1ISUS}$ | $\hat{\theta}_{2ISUS}$ | $\hat{p}_{ISUS}$ | $\hat{\theta}_{1US}$ | $\hat{\theta}_{2US}$ | $\hat{p}_{US}$ |
| | | | | Uniform Priors | | | | | |
| 300 | 0.00443 | 127.92433 | 0.48775 | 0.00566 | 162.56980 | 0.48649 | 0.00478 | 137.38090 | 0.48814 |
| | **0.00000** | **459.91666** | **0.00015** | **0.00000** | **174.23380** | **0.00018** | **0.00000** | **143.73830** | **0.00014** |
| 400 | 0.00459 | 130.14047 | 0.48175 | 0.00495 | 147.82710 | 0.48087 | 0.00479 | 135.86350 | 0.48253 |
| | **0.00000** | **369.77489** | **0.00033** | **0.00000** | **2.38040** | **0.00037** | **0.00000** | **182.42670** | **0.00031** |
| 630 | 0.00476 | 146.94020 | 0.50000 | 0.00485 | 149.80450 | 0.50157 | 0.00486 | 149.87900 | 0.50000 |
| | **0.00000** | **5.90380** | **0.00000** | **0.00000** | **0.18882** | **0.00000** | **0.00000** | **0.25911** | **0.00000** |
| | | | | Jeffreys Priors | | | | | |
| | $\hat{\theta}_{1ML}$ | $\hat{\theta}_{2ML}$ | $\hat{p}_{ML}$ | $\hat{\theta}_{1ISJS}$ | $\hat{\theta}_{2ISJS}$ | $\hat{p}_{ISJS}$ | $\hat{\theta}_{1JS}$ | $\hat{\theta}_{2JS}$ | $\hat{p}_{JS}$ |
| 300 | 0.00443 | 127.92433 | 0.48775 | 0.00768 | 128.65234 | 0.42190 | 0.00463 | 133.76613 | 0.48860 |
| | **0.00000** | **459.91666** | **0.00015** | **0.00001** | **429.22125** | **0.00610** | **0.00000** | **243.48079** | **0.00013** |
| 400 | 0.00459 | 130.14047 | 0.48175 | 0.00459 | 152.50032 | 0.51130 | 0.00467 | 132.95324 | 0.48283 |
| | **0.00000** | **369.77489** | **0.00033** | **0.00000** | **9.79892** | **0.00013** | **0.00000** | **269.51009** | **0.00030** |
| 630 | 0.00476 | 146.94020 | 0.50000 | 0.00476 | 146.47420 | 0.50052 | 0.00476 | 146.94020 | 0.50000 |
| | **0.00000** | **5.90380** | **0.00000** | **0.00000** | **8.38563** | **0.00000** | **0.00000** | **5.90380** | **0.00000** |

Table 7.2: ML Estimates, Bayes Estimates and their Estimated Risks under GELF using Real Life Example with $c = 1.2$.

| | ML Estimates | | | Approximate Bayes Estimates | | | Actual Bayes Estimates | | |
|---|---|---|---|---|---|---|---|---|---|
| | | | | Uniform Priors | | | | | |
| $T$ | $\hat{\theta}_{1ML}$ | $\hat{\theta}_{2ML}$ | $\hat{p}_{ML}$ | $\hat{\theta}_{1ISUG}$ | $\hat{\theta}_{2ISUG}$ | $\hat{p}_{ISUG}$ | $\hat{\theta}_{1UG}$ | $\hat{\theta}_{2UG}$ | $\hat{p}_{UG}$ |
| 300 | 0.00443 | 127.92433 | 0.48775 | 0.00565 | 162.11530 | 0.48556 | 0.00440 | 127.12260 | 0.47506 |
| | **0.00354** | **0.01627** | **0.00044** | **0.02248** | **0.00499** | **0.00061** | **0.00436** | **0.01758** | **0.00185** |
| 400 | 0.00459 | 130.14047 | 0.48175 | 0.00487 | 146.23000 | 0.47506 | 0.00456 | 129.83810 | 0.47450 |
| | **0.00099** | **0.01295** | **0.00098** | **0.00039** | **0.00032** | **0.00185** | **0.00136** | **0.01338** | **0.00193** |
| 630 | 0.00476 | 146.94020 | 0.50000 | 0.00475 | 146.71720 | 0.49633 | 0.00475 | 146.64530 | 0.49455 |
| | **0.00000** | **0.00019** | **0.00000** | **0.00001** | **0.00023** | **0.00004** | **0.00000** | **0.00024** | **0.00009** |
| | | | | Jeffreys Priors | | | | | |
| | $\hat{\theta}_{1ML}$ | $\hat{\theta}_{2ML}$ | $\hat{p}_{ML}$ | $\hat{\theta}_{1ISJG}$ | $\hat{\theta}_{2ISJG}$ | $\hat{p}_{ISJG}$ | $\hat{\theta}_{1JG}$ | $\hat{\theta}_{2JG}$ | $\hat{p}_{JG}$ |
| 300 | 0.00443 | 127.92433 | 0.48775 | 0.00751 | 126.54607 | 0.41972 | 0.00427 | 124.00001 | 0.47592 |
| | **0.00354** | **0.01627** | **0.00044** | **0.18096** | **0.01855** | **0.02059** | **0.00816** | **0.02319** | **0.00172** |
| 400 | 0.00459 | 130.14047 | 0.48175 | 0.00446 | 147.19941 | 0.50466 | 0.00444 | 127.04861 | 0.47487 |
| | **0.00099** | **0.01295** | **0.00098** | **0.00303** | **0.00015** | **0.00006** | **0.00343** | **0.01770** | **0.00188** |
| 630 | 0.00476 | 146.94020 | 0.50000 | 0.00466 | 143.40680 | 0.49489 | 0.00466 | 143.70640 | 0.49455 |
| | **0.00000** | **0.00019** | **0.00000** | **0.00030** | **0.00118** | **0.00008** | **0.00035** | **0.00106** | **0.00009** |

Table 7.3: ML Estimates, Bayes Estimates and their Estimated Risks under GELF using Real Life Example with $c = -1.2$.

| | ML Estimates | | | Approximate Bayes Estimates | | | Actual Bayes Estimates | | |
|---|---|---|---|---|---|---|---|---|---|
| | | | | Uniform Priors | | | | | |
| $T$ | $\hat{\theta}_{1ML}$ | $\hat{\theta}_{2ML}$ | $\hat{p}_{ML}$ | $\hat{\theta}_{1ISUG}$ | $\hat{\theta}_{2ISUG}$ | $\hat{p}_{ISUG}$ | $\hat{\theta}_{1UG}$ | $\hat{\theta}_{2UG}$ | $\hat{p}_{UG}$ |
| 300 | 0.00443 | 127.92433 | 0.48775 | 0.00735 | 168.82486 | 0.46648 | 0.00482 | 138.38481 | 0.48929 |
| | **0.00375** | **0.01842** | **0.00045** | **0.11539** | **0.01028** | **0.00357** | **0.00010** | **0.00433** | **0.00034** |
| 400 | 0.00459 | 130.14047 | 0.48175 | 0.00588 | 127.88259 | 0.45218 | 0.00481 | 136.42813 | 0.48324 |
| | **0.00102** | **0.01446** | **0.00101** | **0.02938** | **0.01850** | **0.00758** | **0.00009** | **0.00613** | **0.00085** |
| 630 | 0.00476 | 146.94023 | 0.50000 | 0.00488 | 149.63829 | 0.50056 | 0.00486 | 150.17186 | 0.50049 |
| | **0.00000** | **0.00019** | **0.00000** | **0.00045** | **0.00000** | **0.00000** | **0.00034** | **0.00002** | **0.00000** |
| | | | | Jeffreys Priors | | | | | |
| | $\hat{\theta}_{1ML}$ | $\hat{\theta}_{2ML}$ | $\hat{p}_{ML}$ | $\hat{\theta}_{1ISJG}$ | $\hat{\theta}_{2ISJG}$ | $\hat{p}_{ISJG}$ | $\hat{\theta}_{1JG}$ | $\hat{\theta}_{2JG}$ | $\hat{p}_{JG}$ |
| 300 | 0.00443 | 127.92433 | 0.48775 | 0.00635 | 154.71462 | 0.43588 | 0.00467 | 134.71840 | 0.48971 |
| | **0.00375** | **0.01842** | **0.00045** | **0.05365** | **0.00088** | **0.01434** | **0.00029** | **0.00800** | **0.00031** |
| 400 | 0.00459 | 130.14047 | 0.48175 | 0.00516 | 143.90598 | 0.45927 | 0.00469 | 133.50591 | 0.48354 |
| | **0.00102** | **0.01446** | **0.00101** | **0.00454** | **0.00101** | **0.00538** | **0.00017** | **0.00950** | **0.00082** |
| 630 | 0.00476 | 146.94023 | 0.50000 | 0.00476 | 147.62833 | 0.50108 | 0.00477 | 147.23303 | 0.50049 |
| | **0.00000** | **0.00019** | **0.00000** | **0.00000** | **0.00010** | **0.00000** | **0.00000** | **0.00015** | **0.00000** |